\documentclass[11pt]{article}
\usepackage{latexsym,amsmath,amscd,amssymb,graphicx,amsbsy,color,xcolor,amsfonts,amsthm}
\newtheorem{defn}{Definition}

\begin{document}

\title{Covariant Uniform Acceleration \thanks{Supported in part by the German-Israel Foundation for Scientific Research and Development: GIF No. 1078-107.14/2009}}
\author{Yaakov Friedman and Tzvi Scarr\\Jerusalem College of Technology\\
Departments of Mathematics and Physics \\
P.O.B. 16031 Jerusalem 91160, Israel\\
e-mail:friedman@jct.ac.il, tzviscarr@gmail.com}
\maketitle

\begin{abstract}
We derive a 4D \emph{covariant} Relativistic Dynamics Equation. This equation canonically extends the 3D relativistic dynamics equation $\mathbf{F}=\frac{d\mathbf{p}}{dt}$, where $\mathbf{F}$ is the 3D force and $\mathbf{p}=m_0\gamma\mathbf{v}$ is the 3D relativistic momentum. The \emph{standard} 4D equation $F=\frac{dp}{d\tau}$ is only partially covariant. To achieve full Lorentz covariance, we replace the four-force $F$ by a rank 2 antisymmetric \emph{tensor} acting on the four-velocity. By taking this tensor to be constant, we obtain a covariant definition of \emph{uniformly accelerated motion}. This solves a problem of Einstein and Planck.

We compute \emph{explicit} solutions for uniformly accelerated motion. The solutions are divided into four \emph{Lorentz-invariant} types: null, linear, rotational, and general.  For \emph{null} acceleration, the worldline is \emph{cubic} in the time. \emph{Linear} acceleration covariantly extends 1D hyperbolic motion, while \emph{rotational} acceleration  covariantly extends pure rotational motion.

We use Generalized Fermi-Walker transport to construct a \emph{uniformly accelerated family} of inertial frames which are instantaneously comoving to a uniformly accelerated observer. We explain the connection between our approach and that of Mashhoon. We show that our solutions of uniformly accelerated motion have constant acceleration in the comoving frame. Assuming the Weak Hypothesis of Locality, we obtain local spacetime transformations from a uniformly accelerated frame $K'$ to an inertial frame $K$. The spacetime transformations between two uniformly accelerated frames with the same acceleration are \emph{Lorentz}. We compute the metric at an arbitrary point of a uniformly accelerated frame.

We obtain velocity and acceleration transformations from a uniformly accelerated system $K'$ to an inertial frame $K$. We introduce the \emph{4D velocity}, an adaptation of Horwitz and Piron's notion of ``off-shell." We derive the general formula for the time dilation between accelerated clocks.  We obtain a formula for the angular velocity of a uniformly accelerated object. Every rest point of $K'$ is uniformly accelerated, and its acceleration is a function of the observer's acceleration and its position. We obtain an interpretation of the Lorentz-Abraham-Dirac equation as an acceleration transformation from $K'$ to $K$.
\end{abstract}

\section{Introduction}\label{Intro}

$\;\;\;$

Newton's Second Law $\mathbf{F}=m\mathbf{a}$ can be written as $\mathbf{F}=\frac{d\mathbf{p}}{dt}$, where $\mathbf{p}=m\mathbf{v}$ is the classical momentum. In special relativity, the classical momentum is replaced by the relativistic momentum $\mathbf{p}=m_0\gamma\mathbf{v}$, and Newton's Second Law is replaced by the standard 3D relativistic dynamics equation \cite{gps,Rindler,Moller}
\begin{equation}\label{dudtg}
\mathbf{F}=m_0\frac{d(\gamma\mathbf{v})}{dt}.
\end{equation}
When the 3D force $\mathbf{F}$ is \emph{constant}, the solutions to (\ref{dudtg}) are traditionally called \emph{uniformly accelerated motion}. Equation (\ref{dudtg}), however, is covariant only with respect to the \emph{little Lorentz group} and \emph{not} covariant with respect to the full Lorentz group.

As a 4D extension of (\ref{dudtg}), we have
\begin{equation}\label{4drde}
F=\frac{dp}{d\tau},
\end{equation}
where $F$ is the four-force, $p$ is the four-momentum, and $\tau$ is proper time (see \cite{gps}). Unfortunately, equation (\ref{4drde}) is also covariant only with respect to the little Lorentz group. Moreover, when $F$ is a constant, as in a homogeneous gravitational field, equation (\ref{4drde}) has no solution! This follows from the fact that the four-velocity and the four-acceleration are perpendicular. This was noticed by Planck, who wrote to Einstein about it. This, in turn, prompted Einstein to submit a ``correction" \cite{E1} to \cite{E2}. In the correction, he states that the ``concept `uniformly accelerated' needs further clarification." This was a call for a fully Lorentz covariant relativistic dynamics equation and for a better definition of ``uniform acceleration."

It was clear, even in 1908, that the \emph{physical} definition of ``uniformly accelerated motion" is motion whose acceleration is \emph{constant in the comoving frame}. This definition is found widely in the literature, as early as \cite{Mink} and \cite{Born2}, again in \cite{LL}, and as recently as \cite{Rohrlich2} and \cite{Lyle}. This definition is natural, since the acceleration in the comoving frame is ``precisely the push we feel when sitting in an accelerating rocket" or automobile. Similarly, ``by the equivalence principle, the gravitational field in our terrestrial lab is the negative of our proper acceleration, our instantaneous rest frame being an imagined Einstein cabin falling with acceleration $g$" (\cite{Rindler}, page 71). Indeed, we will, at times, invoke the Equivalence Principle and interpret accelerations as the effect of a gravitational field of a massive object.

If the acceleration is constant in the comoving frame, then the length of the four-acceleration $a$ is constant:
\begin{equation}\label{a2isconst}
a^\mu a_\mu= \hbox{constant}.
\end{equation}
Equation (\ref{a2isconst}) is a good candidate to replace (\ref{4drde}).  It's even \emph{fully} Lorentz covariant. However, as in the case of equations (\ref{dudtg}) and (\ref{4drde}), existing techniques have produced only \emph{1D hyperbolic motion} as solutions to (\ref{a2isconst}). There are clearly some missing solutions, since equation (\ref{a2isconst}) is covariant, while the class of 1D hyperbolic motions is \emph{not}.

We are thus faced with two problems:
\vskip0.2cm
\begin{itemize}
\item[(1)] Can $\mathbf{F}=\frac{d\mathbf{p}}{dt}$ be extended to a 4D Lorentz covariant version?
\item[(2)] What is the ``right" equation for uniformly accelerated motion?
\end{itemize}
In this paper, we derive a 4D Lorentz covariant Relativistic Dynamics Equation:
\begin{equation}\label{uam1}
c\frac{du^{\mu}}{d\tau}=A^{\mu}_{\nu}u^{\nu},
\end{equation}
where $u$ is the four-velocity, $\tau$ is proper time, and $A_{\mu\nu}$ is a rank 2 antisymmetric tensor, or, equivalently, $A^\mu_\nu$ is skew adjoint with respect to the Minkowski inner product $\eta_{\mu\nu}=\operatorname{diag}(1,-1,-1,-1)$. As will be shown here, equation (\ref{uam1}) has the following advantages:
\begin{itemize}
\item [$\bullet$] It canonically extends the Relativistic Dynamics Equation (\ref{dudtg}) and is covariant with respect to the \emph{full Lorentz group}
\item [$\bullet$] By redefining \emph{uniformly accelerated motion} as the solutions to (\ref{uam1}) when $A$ is \emph{constant}, we obtain the clarification that Einstein was looking for
\item[$\bullet$] It admits four Lorentz-invariant classes of solutions: null acceleration, linear acceleration, rotational acceleration, and general acceleration. The null, rotational, and general classes were previously unknown. The linear class is a \emph{covariant extension} of 1D hyperbolic motion and contains the motion of an object in a \emph{homogeneous gravitational field}
\item[$\bullet$] It can be extended in a straightforward manner to obtain a covariant definition of the ``comoving frame" of a uniformly accelerated observer. In this comoving frame, all of the solutions of (\ref{uam1}) have constant acceleration.
\item[$\bullet$] It can be modified to accommodate a universal maximal acceleration. Thus, this paper is an important step in the study of evidence for and implications of the existence of a universal maximal acceleration (see \cite{FG10,F11Ann,FR}).
\end{itemize}

The plan of the paper is as follows. In section \ref{1dhm}, we derive a 4D Lorentz covariant \emph{Relativistic Dynamics Equation}. We also show that equations (\ref{dudtg}) and (\ref{4drde}) are \emph{not} Lorentz covariant. They are, however, canonically embedded in (\ref{uam1}). By taking the tensor $A$ to be constant, we obtain a covariant definition of \emph{uniformly accelerated motion}. In section \ref{explicitsolns}, we obtain \emph{explicit} solutions to equation (\ref{uam1}) in the case of a constant force. Our solutions are divided into four Lorentz-invariant types: null acceleration, linear acceleration, rotational acceleration and general acceleration. The linear acceleration is a covariant extension of 1D hyperbolic motion. The rotational class is a covariant extension of rotations. By integrating our solutions, we obtain trajectories of a uniformly accelerated observer. We conclude this section by computing the nonrelativistic limits of our solutions.

In section \ref{ccmf}, we attach to our observer a \emph{uniformly accelerated frame}, which is defined covariantly by extending equation (\ref{uam1}). Our technique is essentially that of Generalized Fermi-Walker transport and provides a covariant definition of the \emph{comoving frame of a uniformly accelerated observer}. We then show that all of our solutions to (\ref{uam1}) have \emph{constant acceleration in the comoving frame}. We also show that our definition of the comoving frame is equivalent to that of Mashhoon \cite{Mash3}. We also have the surprising result that if two uniformly accelerated frames have a common acceleration tensor $A$, then the spacetime transformations between them are \emph{Lorentz}, despite the fact that neither frame is inertial. In section \ref{spacetimetrans}, we use the Weak Hypothesis of Locality to compute the spacetime transformations from a uniformly accelerated frame to an inertial frame. We show that these transformations extend the Lorentz transformations. Section \ref{examples} is devoted to examples of these transformations.

In section \ref{veltransandtd}, we adapt Horwitz and Piron's notion of ``off-shell" \cite{offshell} to the four-velocity. We call the new notion the \emph{4D velocity} and use it to derive velocity transformations from $K'$ to $K$. We show that when $K'$ is inertial, our velocity transformations reduce to the usual Einstein velocity addition.  Using the ratio between the four-velocity and the 4D velocity, we obtain, in section \ref{timedilation}, the general formula for the time dilation between clocks located at different positions in $K'$. We also derive a formula for the \emph{angular velocity of a uniformly accelerated body}. In section \ref{arbobs}, we consider an observer located at an arbitrary point (not necessarily the origin, as has been the case until now) of a uniformly accelerated frame and show that this observer is also uniformly accelerated. We then derive the transformation formulas for the components $\mathbf{g}$ and $\boldsymbol{\omega}$ of the acceleration tensor $A$. Section \ref{acctrans} is devoted to acceleration transformations from $K'$ to $K$. We use our acceleration transformations to explain the Lorentz-Abraham-Dirac equation. In section \ref{conc}, we discuss the continuation of this research.

Accelerated reference frames were also studied in \cite{Nelson} and \cite{Turyshev}.

\section{Covariant Relativistic Dynamics Equation}\label{1dhm}
$\;\;\;$

In this section, we derive a 4D Lorentz covariant \emph{Relativistic Dynamics Equation} (equation (\ref{uam2}) below). We also show that equations (\ref{dudtg}) and (\ref{4drde}) are \emph{not} Lorentz covariant. They are, however, canonically embedded in (\ref{uam2}). First, we will review some basic notions and establish our notation.

\subsection{Basic Notions}\label{bn}
$\;\;\;$

In flat Minkowski space, the spacetime coordinates of an event are denoted by
$x^{\mu}\;(\mu=0,1,2,3)$, with $x^0=ct$. The \emph{Minkowski inner product} is
\begin{equation}\label{Minkip}
x\cdot y=\eta_{\mu\nu}x^{\mu}y^{\nu},
\end{equation}
where $\eta$ is the Minkowski metric $\eta_{\mu\nu}=\operatorname{diag}(1,-1,-1,-1)$. The worldline of a particle is $x(t)=(ct,\mathbf{x}(t))$. The particle's 3D velocity is  $\mathbf{v}=\frac{d\mathbf{x}}{dt}$. Then $\frac{dx}{dt}=(c,\mathbf{v})$, and the dimensionless scalar $\gamma$ is defined by
\begin{equation}\label{defgamma}
\gamma=\gamma(\mathbf{v})=\frac{1}{\left|\frac{dx}{cdt}\right|}=\frac{1}{\sqrt{1-\frac{\mathbf{v}^2}{c^2}}}.
\end{equation}
The particle's \textit{proper time}, denoted by $\tau$, is defined by
\begin{equation}\label{defproptime}
\gamma d\tau=dt.
\end{equation}
Since $cd\tau=ds$, where $ds$ is the differential of arc length along the particle's worldline, the proper time is a Lorentz invariant quantity.
The particle's dimensionless \textit{four-velocity} $u^{\mu}$ is defined, as usual, by
\begin{equation}\label{4vel}
u^{\mu}=\frac{dx^{\mu}}{ds}=\frac{1}{c}\frac{dx^{\mu}}{d\tau}, \quad (u^0,u^1,u^2,u^3)=\gamma\left(1,\frac{\mathbf{v}}{c}\right),
\end{equation}
and its \textit{proper velocity} $\mathbf{u}$ is $c$ times the spatial part of the four-velocity:
\begin{equation}\label{propvel}
\mathbf{u}=\gamma(\mathbf{v})\mathbf{v}.
\end{equation}
A straightforward calculation shows that we can write $\gamma$ as a function of the proper velocity:
\begin{equation}\label{gammau}
\gamma=\sqrt{1+\frac{\mathbf{u}^2}{c^2}}.
\end{equation}
The four-velocity always has ``length" $1$ in the Minkowski metric:
\begin{equation}\label{leng}
|u|=\sqrt{\eta_{\mu\nu}u^{\mu}u^{\nu}}=1.
\end{equation}
The particle's \textit{four-acceleration} $a^{\mu}$ is defined by
\begin{equation}\label{4a}
a^{\mu}=c\frac{du^{\mu}}{d\tau}
\end{equation}
and has units of acceleration. Differentiating $u \cdot u = 1$, we see that the four-acceleration and the four-velocity are always perpendicular:
\begin{equation}\label{4perp}
u\cdot a=\eta_{\mu\nu}u^{\mu}a^{\nu}=0.
\end{equation}
This implies that the four-acceleration is spacelike.

The \emph{rest-mass} of an object is denoted by $m_0$, and we let $m=m(\mathbf{v})=m_0\gamma=m_0\gamma(\mathbf{v})$. The \emph{3D momentum} is $\mathbf{p}=m\mathbf{v}=m_0\gamma\mathbf{v}=m_0\mathbf{u}$, and the \emph{four-momentum} is $p=m_0cu=(m_0\gamma c,m_0\gamma\mathbf{v})=(mc,\mathbf{p})$.

\subsection{Embedding $\mathbf{F}=\frac{d\mathbf{p}}{dt}$ in Four Dimensions}\label{4Dcrde}
$\;\;\;$
The standard \emph{Relativistic Dynamics Equation} is the 3D equation
\begin{equation}\label{dudtg2}
\mathbf{F}=\frac{d\mathbf{p}}{dt}.
\end{equation}
In special relativity, however, we require a 4D version of this equation. Since the 3D vector $\mathbf{p}$ is part of the four-momentum $p$, we seek an appropriate expression for $\frac{dp}{d\tau}$ (or $\frac{du}{d\tau}$).

It is natural to consider the 4D equation
\begin{equation}\label{4drde2}
F=\frac{dp}{d\tau}.
\end{equation}
This equation, however, \emph{has no solution} when $F$ is a \emph{constant} four-vector. To see this, suppose $F$ is constant. Then, since $F \sim a$, equation (\ref{4perp}) implies that $a$ is both lightlike and perpendicular to the timelike vector $u$, which is impossible. This implies that the four-acceleration \textit{cannot} be constant in an inertial frame. Hence, equation (\ref{4drde2}) cannot be used to model constant-force motion and is inappropriate as a dynamics equation.

Next, we show that $\mathbf{F}=\frac{d\mathbf{p}}{dt}=m_0\frac{d\mathbf{u}}{dt}$ can be written in the form $c\frac{du}{d\tau}=Au$, where $A$ is an antisymmetric tensor. Since $u=(\gamma,\mathbf{u}/c)$, we have
\begin{equation}\label{m0dudt}
\frac{du}{d\tau}=\left(\frac{d\gamma}{d\tau},\frac{d\mathbf{u}}{cd\tau}\right).
\end{equation}
Using (\ref{gammau}) and then (\ref{defproptime}), we have
\begin{equation}\label{dmdtau}
\frac{d\gamma}{d\tau}=\frac{d\gamma}{d\mathbf{u}}\frac{d\mathbf{u}}{d\tau}=\frac{\mathbf{u}/c^2}{\gamma}\cdot \gamma \frac{d\mathbf{u}}{dt}=\frac{\mathbf{u}}{c^2}\cdot \frac{d\mathbf{u}}{dt}=\frac{1}{m_0c^2}\mathbf{u}\cdot\mathbf{F},
\end{equation}
and
\begin{equation}\label{dpdtau}
\frac{d\mathbf{u}}{d\tau}=\gamma\frac{d\mathbf{u}}{dt}=\frac{\gamma}{m_0}\mathbf{F}.
\end{equation}
Combining (\ref{dmdtau}) and (\ref{dpdtau}), we have
\begin{equation}\label{rdematrix}
c\frac{du}{d\tau}
=\frac{1}{m_0}\left(\begin{array}{ll}0 & \mathbf{F}^T\\ \mathbf{F} & 0 \end{array}  \right)u,
\end{equation}
where the superscript $T$ denotes matrix transposition. This shows that the 3D Relativistic Dynamics Equation (\ref{dudtg2}) is equivalent to
\begin{equation}\label{embiff}
c\frac{du}{d\tau}=Au,\;\;\mbox{ with } \;\;A=\frac{1}{m_0}\left(\begin{array}{ll}0 & \mathbf{F}^T\\ \mathbf{F} & 0 \end{array}  \right).
\end{equation}
Note that $A$ is an antisymmetric tensor of the particular form $\left(\begin{array}{ll}0 & \mathbf{g}^T\\ \mathbf{g} & 0 \end{array}  \right)$, where $\mathbf{g}=\frac{1}{m_0}\mathbf{F}$. As an operator, $A=A^\mu_\nu$ has mixed indices, one upper and one lower. If we lower the upper index using the Minkowski metric, $A_{\mu\nu}$ is antisymmetric.

\subsection{Achieving Lorentz Covariance}\label{canemb}
$\;\;\;$
A tensor of the form $\left(\begin{array}{ll}0 & \mathbf{g}^T\\ \mathbf{g} & 0 \end{array}  \right)$ is not Lorentz covariant. In fact, a Lorentz transformation of such a tensor will, in general, produce \emph{any} antisymmetric tensor. Therefore, in order to achieve Lorentz covariance, we must allow $A$ to be \emph{any} antisymmetric tensor. In fact, if $F=Au$, where is a tensor, then $A$ \emph{must} be antisymmetric. To see this, first note that since, from (\ref{4drde2}), $F\sim a$, we have, from (\ref{4perp}), that $u \cdot F = 0$. Substituting $F=Au$, we obtain
\[ 0
=\eta_{\mu\nu}u^{\mu}F^{\nu}=\eta_{\mu\nu}u^{\mu}A^{\nu}_{\alpha}u^{\alpha}
=u^{\mu}A_{\mu\alpha} u^{\alpha},\] and so
\begin{equation}\label{antisy}
A_{\alpha\beta}=-A_{\beta\alpha}.
\end{equation}
The need for antisymmetry can be understood as follows. The Lorentz transformation produces a \emph{rotation} in Minkowski spacetime. Similarly, \textit{acceleration} can be interpreted as a rotation of the four-velocity, since the four-acceleration is perpendicular to the four-velocity. It is known that a rotation in 3D Euclidean space is given by the exponent of an antisymmetric tensor. The antisymmetry of $A$ is the 4D extension of this fact.

We thus arrive at a
\emph{4D Lorentz covariant Relativistic Dynamics Equation}
\begin{equation}\label{uam2}
\boxed{c\frac{du^{\mu}}{d\tau}=A^{\mu}_{\nu}u^{\nu}}
\end{equation}
\vskip0.2cm\noindent
where $A_{\mu\nu}$ is an antisymmetric rank 2 tensor, or, equivalently, $A^\mu_\nu$ is skew adjoint with respect to the inner product (\ref{Minkip}). The components of $A$ have units of acceleration and may be \emph{functions} of the position $x$ and the four-velocity $u$. We refer to $A$ as the \emph{acceleration tensor} associated with the given motion. Equation (\ref{uam2}) solves the first problem mentioned in the introduction.

In the $1+3$ decomposition, the tensors $A_{\mu\nu}$ and $A_\nu^\mu$ take the form
\begin{equation}\label{aab}
A_{\mu\nu}(\mathbf{g},\boldsymbol{\omega})=\left(\begin{array}{cc}0 & \mathbf{g}^T\\ &
\\-\mathbf{g}&-c\pi(\boldsymbol{\omega})\end{array}\right),\quad
A_\nu^\mu(\mathbf{g},\boldsymbol{\omega})=\left(\begin{array}{cc}0 & \mathbf{g}^T\\ &
\\\mathbf{g}&c\pi(\boldsymbol{\omega})\end{array}\right),
\end{equation}
where $\mathbf{g}$ is a 3D vector with units of acceleration, $\boldsymbol{\omega}$ is a 3D vector with units of $1/\hbox{time}$, and, for any 3D vector $\boldsymbol{\omega}=(\omega^1,\omega^2,\omega^3)$,
\[ \pi(\boldsymbol{\omega})= \varepsilon_{ijk}\omega^k, \]
where $\varepsilon_{ijk}$ is the Levi-Civita tensor. The factor $c$ in $A$ provides the necessary units of acceleration. The 3D vectors $\mathbf{g}$ and $\boldsymbol{\omega}$ are related to the linear, or translational, acceleration and the angular velocity, respectively, of the motion. We will obtain a more precise explanation of the physical meaning of these vectors in sections \ref{ccmf} and \ref{timedilation}. We will also show there that if a uniformly accelerated system was at rest at time $t=0$, then $\mathbf{F}/m_0$ is the constant acceleration in the comoving frame.

By (\ref{embiff}), $\mathbf{F}=\frac{d\mathbf{p}}{dt}$ is equivalent to a 4D equation
\begin{equation}\label{auspat}
c\frac{du}{d\tau}=\left(\begin{array}{cc}0 & \mathbf{g}^T\\ &
\\\mathbf{g}&c\pi(\boldsymbol{\omega})\end{array}\right)u,
\end{equation}
where $\boldsymbol{\omega}=0$. Hence, $\mathbf{F}=\frac{d\mathbf{p}}{dt}$ is covariant
only with respect to transformations which preserve the condition $\boldsymbol{\omega}=0$. A straightforward calculation shows that the only Lorentz transformations which preserve the condition $\boldsymbol{\omega}=0$ are boosts in the direction of $\mathbf{F}$ and spatial rotations about the axis of $\mathbf{F}$ . The little Lorentz group, as defined in \cite{Wigner}, is the stabilizer of the spatial axis in a given direction, which we may choose to be the direction of the force $\mathbf{F}$. Thus, $\mathbf{F}=\frac{d\mathbf{p}}{dt}$ is covariant only with respect to this little Lorentz group and \emph{not} to the full Lorentz group. It follows immediately that \emph{1D hyperbolic motion} is also covariant only with respect to this little Lorentz group.

For an additional proof, note that the spatial part of (\ref{auspat}) is
\begin{equation}\label{3dpart}
\frac{d\mathbf{u}}{d\tau} = \gamma\mathbf{g} + \mathbf{u}\times \boldsymbol{\omega},\;\;\hbox{or}\;\; \frac{d\mathbf{p}}{dt} = \mathbf{F} + \gamma^{-1}\mathbf{p}\times \boldsymbol{\omega}.
\end{equation}
Hence, $\mathbf{F}=\frac{d\mathbf{p}}{dt}$ if and only if $\boldsymbol{\omega}=0$.

By taking the tensor $A$ to have \emph{constant} components, we obtain a covariant definition of uniformly accelerated motion, thus solving the second problem mentioned in the introduction. We define \emph{uniformly accelerated motion} as motion whose four-velocity $u(\tau)$ is a
solution to the initial value problem
\begin{equation}\label{uam2ivp}
c\frac{du^{\mu}}{d\tau}=A^{\mu}_{\nu}u^{\nu}\quad , \quad u(0)=u_0,
\end{equation}
where $A_{\mu\nu}$ is an antisymmetric rank 2 tensor with \emph{constant} components, or, equivalently, $A^\mu_\nu$ is constant and skew adjoint with respect to the inner product (\ref{Minkip}). In the next section, we compute \emph{explicit} solutions to (\ref{uam2ivp}).

\section{Explicit Trajectories for Uniformly Accelerated Motion}\label{explicitsolns}

$\;\;\;$
In this section, we obtain explicit trajectories for uniformly accelerated motion, that is, we obtain the explicit solutions $u(\tau)$ of (\ref{uam2ivp}). It is known (see \cite{CO}, page \textbf{1}-65) that for a given initial condition $u(0)=u_0$, (\ref{uam2ivp}) has the unique solution
\begin{equation}\label{exponent solution}
    u(\tau)=\exp (A\tau/c)u_0=\left( \sum_{n=0}^{\infty}\frac{A^n}{n!c^n}\tau ^n\right)u_0\,.
\end{equation}
The worldline $\widehat{x}(\tau)$ of a uniformly accelerated observer may then be obtained by integrating $u(\tau)$.

Since $A$ is antisymmetric, all solutions of the form (\ref{exponent solution}) are Lorentz transformations of the initial velocity $u_0$, with an angle that is linear in $\tau$.

\subsection{Lorentz-Invariant Classification of Solutions}\label{invclass}

Fix $A=A^\mu_\nu(\mathbf{g},\boldsymbol{\omega})$ as in (\ref{aab}). It can be shown by direct calculation that
\begin{equation}\label{detaminuslambdai}
\det(A-a I)=a^4-l_1a^2-c^2(l_2)^2,
\end{equation}
where $l_1=\mathbf{g}^2-c^2\boldsymbol{\omega}^2$ and $l_2=\mathbf{g}\cdot\boldsymbol{\omega}$ are Lorentz invariants, similar to the two known Lorentz invariants associated with the electromagnetic field. Hence,
the matrix $A^\mu_\nu(\mathbf{g},\boldsymbol{\omega})$ (one upper index, one lower index) has the eigenvalues $\pm\alpha$ and $\pm i\beta$, where
\[ \alpha=\sqrt{\frac{\sqrt{(l_1)^2+4(cl_2)^2}+l_1}{2}}\quad\hbox{ and }\quad \beta=\sqrt{\frac{\sqrt{(l_1)^2+4(cl_2)^2}-l_1}{2}}.\]

We classify our solutions for uniformly accelerated motion into four types, depending on the values of the Lorentz invariants and the eigenvalues of $A^\mu_\nu(\mathbf{g},\boldsymbol{\omega})$:
\[ \begin{array}{ccccc}
\hbox{\underline{Type}} & \; &\hbox{\underline{Lorentz invariants}} &\;& \hbox{\underline{Eigenvalues}} \\
   \; &\;    &\;&\;&\;\\
 \hbox{Null} &\;& l_1=l_2=0 & \; & \alpha=0,\beta=0\\
     \;& \; &\;&\;&\;\\
\hbox{Linear}& \;&l_1>0, l_2=0 & \; & \alpha = \sqrt{l_1} > 0,\beta=0\\
  \; &\;&\;&\;&\;\\
 \hbox{Rotational}& \;&  l_1<0, l_2=0&\; & \alpha=0,\beta =\sqrt{-l_1} > 0\\
     \; &\;&\;&\;&\;\\
 \hbox{General}&\;& l_2\neq 0 &\; & \alpha > 0,\beta > 0
      \end{array}   \]
Note that each type is a \emph{Lorentz-invariant subset}.
\vskip0.3cm
For each of the four types of uniformly accelerated motion, we now obtain the explicit solutions $u(\tau)$ of (\ref{uam2ivp}).

\subsection{Null Acceleration\quad ($\alpha=0,\beta=0$)}

In this case, $|\mathbf{g}|=c|\boldsymbol{\omega}|$ and $\mathbf{g}\perp \boldsymbol{\omega}$. Direct calculation shows that $A^3=0$. Thus, from (\ref{exponent solution}), we have
\begin{equation}\label{utype0}
\boxed{u(\tau)=u(0)+Au(0)\tau/c+\frac{1}{2}A^2u(0)\tau^2/c^2}
\end{equation}
The four-acceleration is
\begin{equation}\label{atype0}
a(\tau)=c\frac{du}{d\tau}= Au(0) + A^2u(0)\tau/c.
\end{equation}
Despite the apparent dependence of the four-acceleration on $\tau$, we will show in section \ref{examples} that the length of $a(\tau)$ is, in fact, constant.

The worldline $\widehat{x}(\tau)$ of a uniformly accelerated observer is, then, \emph{cubic} in $\tau$ and may be obtained by integrating $u(\tau)$:
\begin{equation}\label{xtype0}
\widehat{x}(\tau)= \widehat{x}(0)+c\int_0^{\tau}u(s)ds=\widehat{x}(0) +u(0)c\tau+ \frac{1}{2}Au(0)\tau^2+\frac{1}{6}A^2u(0)\tau^3/c.
\end{equation}
A similar cubic equation was obtained in \cite{Semon} and page 83 of \cite{F04}.

Note that, in general, $A^2\neq 0$, as can be seen from the example
\[A=\left( \begin{array}{rrrr}
0 & 1 & 0 & 0\\1 & 0 & 0 & -1\\0&0&0&0\\0&1&0&0 \end{array}\right)\quad,\quad A^2=\left( \begin{array}{rrrr}
1 & 0 & 0 & -1\\0 & 0 & 0 & 0\\0&0&0&0\\1&0&0&-1 \end{array}\right) .\]

\subsection{Linear, Rotational and General Acceleration}
$\;$
We will obtain the solutions for the remaining three types using the eigenvalues and eigenvectors of $A$. The following two claims use the \emph{linear} extension $\eta^*$ of the Minkowski inner product (\ref{Minkip}) to complex Minkowski space (see \cite{Barut}, p. 12).
\vskip0.3cm\noindent
\textbf{Claim 1}\quad Let $A$ be a skew adjoint matrix with respect to $\eta^*$. Let $a$ be a non-zero eigenvalue of $A$. If $v$ is an eigenvector of $A$ corresponding to $a$, then $v$ is lightlike.
\vskip0.3cm\noindent
To see this, note that
\[ a v^2 = v \cdot a v = v \cdot Av = -Av \cdot v = -a v \cdot v = -a v^2.\]
Since $a \neq 0$, we must have $v^2=0$.  This proves the claim.
\vskip0.3cm\noindent
\textbf{Claim 2}\quad Let $A$ be a skew adjoint matrix with respect to $\eta^*$. If $v_1$ is an eigenvector corresponding to the eigenvalue $a$, and $v_2$ is an eigenvector corresponding to the eigenvalue $\mu$, where $a \neq -\mu$, then $v_1$ and $v_2$ are orthogonal.
\vskip0.3cm\noindent
To see this, note that
\[ a (v_1 \cdot v_2)=a v_1 \cdot v_2 = A v_1 \cdot  v_2 = v_1 \cdot -A v_2 =v_1 \cdot -\mu v_2 = -\mu (v_1 \cdot v_2). \]
Since $a \neq -\mu$, we must have $v_1 \cdot v_2 =0$. This proves the claim.
\vskip0.3cm\noindent
We will also need the following notation. Let $r$ be a real four-vector, and let $\{E_0,E_1,E_2,E_3\}$ be linearly independent (possibly complex) four-vectors. Let $c_0,c_1,c_2,c_3$ be the unique (possibly complex) numbers such that $r=\sum_{k=0}^{3}c_kE_k$. Let $\Im(z)$ denote the imaginary part of a complex scalar or vector $z$. Define
\begin{equation}\label{defDv}
\begin{array}{cc}
D_0(r)=c_0E_0+c_1E_1,& D_1(r)=c_0E_0-c_1E_1, \\
\;&\;\\
D_2(r)=c_2E_2+c_3E_3, & D_3(r)=-2\Im(c_2E_2). \end{array}
\end{equation}
\subsection{Linear Acceleration\quad ($\alpha>0,\beta=0$)}

In this case, we have $\alpha = \sqrt{\mathbf{g}^2-c^2\boldsymbol{\omega}^2}$ and $\mathbf{g}\perp \boldsymbol{\omega}$.
Let $E_0$ and $E_1$ be eigenvectors corresponding to the eigenvalues $\alpha$ and $-\alpha$, respectively, and let $E_2$ and $E_3$ be linearly independent eigenvectors corresponding to the eigenvalue 0. Note that we may choose all of the eigenvectors to be real. Then
\[ u(\tau)=c_0E_0e^{\alpha\tau/c}+ c_1E_1e^{-\alpha\tau/c}+ c_2E_2 + c_3E_3, \]
where $u(0)=\sum_{k=0}^{3}c_kE_k$. Using (\ref{defDv}), with $D_\mu=D_\mu(u(0))$, we have
\begin{equation}\label{utype1}
\boxed{u(\tau)=D_0\cosh(\alpha\tau/c)+D_1\sinh(\alpha\tau/c)+D_2}
\end{equation}
We claim that $D_0,D_1,D_2$ are mutually orthogonal. By Claim 2, $D_2 \in \operatorname{Span}(E_2,E_3)$ is orthogonal to both $D_0$ and $D_1$, which belong to $\operatorname{Span}(E_0,E_1)$. To show that $D_0\cdot D_1 =0$, note that $E_0=(1/2c_0)(D_0+D_1)$ and $E_1=(1/2c_1)(D_0-D_1)$ are lightlike, by Claim 1. Hence,
$(D_0)^2 \pm 2D_0\cdot D_1 + (D_1)^2=0$, implying that $D_0\cdot D_1 =0$. This also implies that $(D_1)^2=-(D_0)^2$.

From (\ref{4a}) and (\ref{utype1}), we obtain
\begin{equation}\label{atype1}
a(\tau)=\alpha D_0\sinh(\alpha\tau/c)+\alpha D_1\cosh(\alpha\tau/c).
\end{equation}
The length of $a(\tau)$ is constant:
\begin{equation}\label{a2type1}
a^2(\tau)=-\alpha^2 (D_0)^2,
\end{equation}
and thus satisfies (\ref{a2isconst}).
Since $a$ is spacelike, $D_0$ is timelike.

The interpretation of the solutions (\ref{utype1}) are as follows. In the plane generated by $D_0$ and $D_1$, there is 1D hyperbolic motion, and this plane is moving in a normal direction with four-velocity $D_2$. These solutions form a \emph{covariant extension of 1D hyperbolic motion}.

\subsection{Rotational Acceleration\quad ($\alpha=0,\beta>0$)}

In this case, we have $\beta = \sqrt{c^2\boldsymbol{\omega}^2-\mathbf{g}^2}$ and $\mathbf{g}\perp \boldsymbol{\omega}$.
Let $E_2$ and $E_3$ be eigenvectors corresponding to the eigenvalues $i\beta$ and $-i\beta$, respectively, and let $E_0$ and $E_1$ be linearly independent eigenvectors corresponding to the eigenvalue 0.  Since the two complex eigenvalues are complex conjugates of each other, we may choose
$E_3=\overline{E}_2$. Then
\[ u(\tau)=c_0E_0+ c_1E_1+ c_2E_2e^{i\beta\tau/c}+ c_3\overline{E}_2 e^{-i\beta\tau/c}, \]
where $u(0)=\sum_{k=0}^{3}c_kE_k$. Since $u(\tau)$ is real, we must have $c_3=\overline{c_2}$. Using (\ref{defDv}), with $D_\mu=D_\mu(u(0))$, we have
\begin{equation}\label{utype2}
\boxed{u(\tau)=D_0+D_2\cos(\beta\tau/c)+D_3\sin(\beta\tau/c)}
\end{equation}

We claim that $D_0,D_2,D_3$ are mutually orthogonal. By Claim 2, $D_0 \in \operatorname{Span}(E_0,E_1)$ is orthogonal to both $D_2$ and $D_3$, which belong to $\operatorname{Span}(E_2,E_3)$.
To show that $D_2\cdot D_3 =0$, note that $E_2=1/2ic_2(iD_2+D_3),E_3=1/2ic_3(iD_2-D_3)$ are lightlike, by Claim 1. Hence,
$-(D_2)^2 \pm 2iD_2\cdot D_3 + (D_3)^2=0$, implying that $D_2\cdot D_3 =0$. This also implies that $(D_2)^2=(D_3)^2$.

From (\ref{4a}) and (\ref{utype2}), we obtain
\begin{equation}\label{atype2}
a(\tau)=-\beta D_2\sin(\beta\tau/c)+\beta D_3\cos(\beta\tau/c).
\end{equation}
The length of $a(\tau)$ is constant:
\begin{equation}\label{a2type2}
a^2(\tau)=\beta^2 (D_2)^2,
\end{equation}
and thus satisfies (\ref{a2isconst}).
Since $a$ is spacelike, $D_2$ is also spacelike. Substituting $\tau=0$ into (\ref{utype2}), we obtain that $D_0$ is timelike.

The interpretation of the solutions (\ref{utype2}) are as follows. In the plane generated by $D_2$ and $D_3$, there is pure rotational motion, and this plane is moving in a normal direction with four-velocity $D_0$. The solutions (\ref{utype2}) form a \emph{covariant extension of pure rotational motion}.

\subsection{General Acceleration\quad ($\alpha>0,\beta>0$)}

Since the four eigenvalues $\pm\alpha,\pm i\beta$ are distinct, there are linearly independent eigenvectors $E_0,E_1,E_2,E_3$ of $\alpha,-\alpha,i\beta,-i\beta$, respectively. Since the two complex eigenvalues are complex conjugates of each other, we may choose
$E_3=\overline{E}_2$. Then
\[ u(\tau)=c_0E_0e^{\alpha\tau/c}+ c_1E_1e^{-\alpha\tau/c}+ c_2E_2e^{i\beta\tau/c}+ c_3\overline{E}_2 e^{-i\beta\tau/c}, \]
where $u(0)=\sum_{k=0}^{3}c_kE_k$.
Since $u(\tau)$ is real, we must have $c_3=\overline{c_2}$. Using (\ref{defDv}), with $D_\mu=D_\mu(u(0))$, we have
\begin{equation}\label{utype4}
\boxed{u(\tau)=D_0\cosh(\alpha\tau/c)+D_1\sinh(\alpha\tau/c)+D_2\cos(\beta\tau/c)+D_3\sin(\beta\tau/c)}
\end{equation}

As in the previous cases, the vectors $D_0,D_1,D_2,D_3$ are mutually orthogonal, $(D_1)^2=-(D_0)^2$, and $(D_2)^2=(D_3)^2$.

From (\ref{4a}) and (\ref{utype4}), we obtain
\begin{equation}\label{t2a}
a(\tau)=\alpha D_0\sinh(\alpha\tau/c)+\alpha D_1\cosh(\alpha\tau/c)-\beta D_2\sin(\beta\tau/c)+\beta D_3\cos(\beta\tau/c).
\end{equation}
The length of $a(\tau)$ is constant:
\begin{equation}\label{t2a2}
a^2(\tau)=-\alpha^2 D_0^2 + \beta^2 D_2^2,
\end{equation}
and thus satisfies (\ref{a2isconst}).

Since, in this case, $\mathbf{g}$ and $\boldsymbol{\omega}$ are not perpendicular, there exists a basis in which they are parallel (see \cite{LL}). Here, we have in fact obtained the explicit form of this basis, namely, $\{D_0,D_1,D_2,D_3\}$. In the plane generated by $D_2$ and $D_3$, there is pure rotational motion, and this plane is uniformly accelerated in a normal direction.

\vskip0.5cm\noindent
This completes all of the cases.
The general solution to (\ref{uam2ivp}) is
\begin{equation}\label{gensolnbytype2}
u(\tau)=\left\{\begin{array}{l}
  u(0)+Au(0)\tau/c+\frac{1}{2}A^2u(0)\tau^2/c^2\;,\;\hbox{if }\alpha=0,\beta=0 \hbox{ (null acceleration)}\\
  \;\\
 D_0\cosh(\alpha\tau/c)+D_1\sinh(\alpha\tau/c)+D_2 \;,\;\hbox{if }\alpha>0,\beta=0 \hbox{ (linear acceleration)}\\
 \;\\
 D_0+D_2\cos(\beta\tau/c)+D_3\sin(\beta\tau/c)\;,\;\hbox{if }\alpha=0,\beta>0 \hbox{ (rotational acceleration)} \\
 \;\\
 D_0\cosh(\alpha\tau/c)+D_1\sinh(\alpha\tau/c)\\
+ D_2\cos(\beta\tau/c)+D_3\sin(\beta\tau/c)\;,\;\hbox{if }\alpha>0,\beta>0 \hbox{ (general acceleration)}
 \end{array} \right\}.
\end{equation}

\subsection{Nonrelativistic Limit}\label{nrl}
$\;\;\;$
Here we compute the nonrelativistic limit ($c \rightarrow \infty$) for uniformly accelerated motion. Notice from (\ref{aab}) that $c$ appears in the definition of the acceleration tensor $A$. However, this is only to provide the correct units. Thus, when taking the limit $c \rightarrow \infty$, we may consider $c\pi(\boldsymbol{\omega})$ as a \emph{constant} $\pi(\boldsymbol{\omega}')$. We refer to this method as ``holding the tensor $A$ constant". This is equivalent to holding the eigenvalues $\alpha$ and $\beta$ constant. Alternatively, when taking the limit $c \rightarrow \infty$, we may ``hold the \emph{components} $\mathbf{g}$ and $\boldsymbol{\omega}$ constant" and let the $c$ of $c\pi(\boldsymbol{\omega})$ also go to infinity.

For null acceleration, we must hold the tensor $A$ constant since letting the $c$ of $c\pi(\boldsymbol{\omega})$ go to infinity breaks the condition $\mathbf{g}^2=c^2\boldsymbol{\omega}^2$. Note that the spatial part of $cu$ is the velocity $\mathbf{v}$ in the nonrelativistic limit, and, in particular, the spatial part of $cu(0)$ is $\mathbf{v}(0)$ in this limit. Thus, the nonrelativistic limit of (\ref{utype0}) is
\begin{equation}\label{classicallimitnull}
\mathbf{v}(t)=\lim_{c\rightarrow\infty}\left(\mathbf{v}(0)+A\mathbf{v}(0)\tau/c +O(c^{-2})\right)=\mathbf{v}(0),
\end{equation}
implying that the velocity in such motion is constant. Thus, the nonrelativistic limit of null acceleration is \emph{zero} acceleration. This justifies the name \emph{null} acceleration.

Next, we compute the nonrelativistic limits for linear, rotational, and general acceleration when we hold the tensor $A$ constant. First, we show that the 3D acceleration $\mathbf{a}=\frac{d\mathbf{v}}{dt}$ is the nonrelativistic limit of the spatial part of $a(\tau)$. To see this, note that the spatial part of $a=c\frac{du}{d\tau}$ is $\frac{d\mathbf{u}}{d\tau}$. Using (\ref{dmdtau}), we have
\begin{equation}\label{spparta}
\frac{d\mathbf{u}}{d\tau}=\frac{d}{d\tau}(\gamma\mathbf{v})=\mathbf{v}\frac{d\gamma}{d\tau}+\gamma\frac{d\mathbf{v}}{d\tau}=\left(\frac{\mathbf{u}}{c^2}\cdot \frac{d\mathbf{u}}{dt}\right)\mathbf{v}+\gamma^2\frac{d\mathbf{v}}{dt},
\end{equation}
which tends to $\frac{d\mathbf{v}}{dt}$ in the nonrelativistic limit.

In the linear case, from (\ref{atype1}), we have
\begin{equation}\label{classicallimitlinear}
\lim_{c  \rightarrow \infty}a(\tau)=\alpha D_1.
\end{equation}
Similarly, for the rotational case, from (\ref{atype2}) we get $\lim_{c  \rightarrow \infty}a(\tau)=\beta D_3$, and for general acceleration, from (\ref{t2a}) we get $\lim_{c  \rightarrow \infty}a(\tau)=\alpha D_1+\beta D_3$. Since $D_1$ and $D_3$ do not depend on $c$, the 3D acceleration $\mathbf{a}$ is constant in all of these cases in the nonrelativistic limit.

To identify the acceleration, we take the nonrelativistic limit of the first equation of (\ref{3dpart}) and substitute $t=0$. This gives  $\mathbf{a}=\mathbf{g} + \mathbf{v}(0)\times \boldsymbol{\omega}$. Thus, the 3D velocity in the nonrelativistic limit  is
\begin{equation}\label{classicallimitlinear2}
\mathbf{v}(t)=\mathbf{v}(0)+(\mathbf{g} + \mathbf{v}(0)\times \boldsymbol{\omega})t.
\end{equation}
The nonrelativistic limit is constant \emph{linear} acceleration. For $\boldsymbol{\omega}=0$, we obtain the usual nonrelativistic result. A nonzero value of $\boldsymbol{\omega}$, however, will change both the magnitude and the direction of the acceleration.

Finally, we compute the nonrelativistic limit by holding the \emph{components} $\mathbf{g}$ and $\boldsymbol{\omega}$ constant. For linear acceleration, we have $\mathbf{g}^2>c^2\boldsymbol{\omega}^2$, and so we cannot let $c \rightarrow\infty$. Thus, we will consider only rotational and general acceleration. Since, in this case, $\alpha,\beta$ and the $D_i$ all vary with $c$, it is easier to pass to the nonrelativistic limit of the first equation of (\ref{3dpart}), which is
\begin{equation}\label{nrl25}
\frac{d\mathbf{v}}{dt} = \mathbf{g} + \mathbf{v}\times \boldsymbol{\omega}.
\end{equation}
Equation (\ref{nrl25}) describes nonrelativistic motion under a Lorentz-type force. If $\mathbf{g}=0$, we obtain rotation with uniform angular velocity $\boldsymbol{\omega}$. If $\boldsymbol{\omega}=0$, we obtain linear acceleration.

\section{Covariant Comoving Frame}\label{ccmf}
$\;\;\;$

In this section, we use Generalized Fermi-Walker transport to define the notion of the \emph{comoving frame of a uniformly accelerated observer}. We then show that in this comoving frame, all of our solutions to equation (\ref{uam2}) have \emph{constant acceleration}. We show that our definition of the comoving frame is equivalent to that of Mashhoon \cite{Mash3}. We also show that if two uniformly accelerated frames have a common acceleration tensor $A$, then the spacetime transformations between them are \emph{Lorentz}, despite the fact that neither frame is inertial.

Gupta and Padmanabhan \cite{gupta} applied Fermi-Walker transport to an accelerated charge and obtained the Lorentz-Abraham-Dirac equation.

\subsection{Uniformly Accelerated Family of Inertial Frames}\label{defcmf}
$\;\;\;$

First, we define the notion of a one-parameter family of inertial frames which are instantaneously comoving to a uniformly accelerated observer. We call such a family a \emph{uniformly accelerated family}. The coordinates in this family of comoving frames will be used as a bridge between the observer's coordinates and the coordinates in an inertial frame $K$, which, without loss of generality, we may take to be the initial frame $K_0$ of the observer himself. The family of frames is constructed by Generalized Fermi-Walker transport of the initial frame $K_0$ along the worldline of the observer.  In the case of 1D hyperbolic motion, this construction reduces to Fermi-Walker transport \cite{H,Hehl}.

In fact, Fermi-Walker transport may \emph{only} be used in the case of 1D hyperbolic motion. This is because Fermi-Walker transport uses only a part of the Lorentz group - the boosts. This subset of the group, however, is not a \emph{subgroup}, since the combination of two boosts entails a rotation. Generalized Fermi-Walker transport, on the other hand, uses the full homogeneous Lorentz group, and can be used for all four types of uniform acceleration: null, linear, rotational, and general.

The construction of the uniformly accelerated family $\{K_\tau:\tau \ge 0\}$ is according to the following definition.
\begin{defn}\label{cmdefeqn}
Let $\widehat{x}(\tau)$ be the worldline of a uniformly accelerated observer whose motion is determined by the acceleration tensor $A$, the initial four-velocity $u(0)$, and the initial position $\widehat{x}(0)$.

To specify the \emph{initial frame $K_0$}, we take the origin at time $\tau=0$ to be $\widehat{x}(0)$. For the basis of $K_0$, choose any orthonormal basis $\widehat{\lambda}=\{u(0),\widehat{\lambda}_{(1)},\widehat{\lambda}_{(2)},\widehat{\lambda}_{(3)}\}$.

For each $\tau>0$, define $K_{\tau}$ as follows. The origin of $K_{\tau}$ at time $\tau$ is set as $\widehat{x}(\tau)$. The basis of $K_{\tau}$ is defined to be the unique solution
$\lambda(\tau)=\{\lambda_{(\kappa)}(\tau):\kappa=0,1,2,3\}$, to the initial
value problem
\begin{equation}\label{uamlam2}
c\frac{d\lambda_{(\kappa)}^{\mu}}{d\tau}=A^{\mu}_{\nu}\lambda_{(\kappa)}^{\nu}\quad
, \quad \lambda_{(\kappa)}(0)=\widehat{\lambda}_{(\kappa)}.
\end{equation}
\end{defn}

We make the following observations about definition \ref{cmdefeqn}:
\begin{itemize}
\item[(1)] The choice of the initial four-velocity $u(0)$ for $\widehat{\lambda}_{(0)}$ is deliberate and required by Generalized Fermi-Walker transport.
\item[(2)] For all $\tau$, we have $\lambda_{(0)}(\tau)=u(\tau)$. This follows immediately from (\ref{uam2ivp}).
\item[(3)] From (\ref{exponent solution}), the unique solution to (\ref{uamlam2}) is
\begin{equation}\label{lambdasoln}
\lambda_{(\kappa)}(\tau)=\exp(A\tau/c)\widehat{\lambda}_{(\kappa)}.
\end{equation}
Using matrix multiplication, we combine the solutions for $\kappa=0,1,2,3$ into one equation
\begin{equation}\label{lambdasolncombined}
\lambda(\tau)=\exp(A\tau/c)\widehat{\lambda}.
\end{equation}
\item[(4)] Since $A$ is antisymmetric, $\exp(A\tau/c)$ is an \emph{isometry}. Thus, $\lambda(\tau)$ is an orthonormal basis.
\item[(5)] Analogously to (\ref{gensolnbytype2}), the general solution to (\ref{uamlam2}) is
\begin{equation}\label{gensolncmf}
\lambda_{(\kappa)}(\tau)=\left\{\begin{array}{l}
  \widehat{\lambda}_{(\kappa)}+A\widehat{\lambda}_{(\kappa)}\tau/c+\frac{1}{2}A^2\widehat{\lambda}_{(\kappa)}\tau^2/c^2\;,\;\hbox{if }\alpha=0,\beta=0 \hbox{ (null acceleration)}\\
  \;\\
 D_0(\widehat{\lambda}_{(\kappa)})\cosh(\alpha\tau/c)+D_1(\widehat{\lambda}_{(\kappa)})\sinh(\alpha\tau/c)+D_2(\widehat{\lambda}_{(\kappa)}),\\
 \quad\quad\quad\quad\hbox{if }\alpha>0,\beta=0 \hbox{ (linear acceleration)}\\
 \;\\
 D_0(\widehat{\lambda}_{(\kappa)})+D_2(\widehat{\lambda}_{(\kappa)})\cos(\beta\tau/c)+D_3(\widehat{\lambda}_{(\kappa)})\sin(\beta\tau/c),\\
 \quad\quad\quad\quad\hbox{if }\alpha=0,\beta>0 \hbox{ (rotational acceleration)} \\
 \;\\
 D_0(\widehat{\lambda}_{(\kappa)})\cosh(\alpha\tau/c)+D_1(\widehat{\lambda}_{(\kappa)})\sinh(\alpha\tau/c)\\
+ D_2(\widehat{\lambda}_{(\kappa)})\cos(\beta\tau/c)+D_3(\widehat{\lambda}_{(\kappa)})\sin(\beta\tau/c)\;,\;\hbox{if }\alpha>0,\beta>0 \hbox{ (general acceleration)}
 \end{array} \right\}.
\end{equation}
\item[(6)] For a given $A$, all four solutions $\lambda_{(\kappa)}(\tau), \kappa=0,1,2,3$ are of the same type (null, linear, rotational, or general).
The comoving frame of a rotating observer ``rotates" along with him, and he feels that in \emph{this} frame, the acceleration is constant.
\item[(7)] Along the worldline, the components of the acceleration tensor $A$ remain constant. To see this, let $A$ denote the tensor as computed in the lab frame $K$, and let $A(\tau)$ denote the tensor as computed in the comoving frame $K_\tau$. Then, since $\lambda(\tau)$ is the change of matrix basis from $K$ to $K_\tau$, we have
\begin{equation}\label{alla}
A(\tau)=\lambda(\tau)^{-1}A\lambda(\tau)=(\exp(A\tau/c)\widehat{\lambda})^{-1}A\exp(A\tau/c)\widehat{\lambda}=
\widehat{\lambda}^{-1} A \widehat{\lambda}=A(0).
\end{equation}
\item[(8)] For all $\tau$, we have $\lambda(\tau)A(\tau)=A\lambda(\tau)$.
\end{itemize}

\subsection{Uniformly Accelerated Frame}\label{uaframe}
$\;$

Two frames are said to be \emph{comoving} at time $\tau$ if at this time, the origins of the two frames coincide, their respective axes are parallel, and they have the same four-velocity.

We now define the notion of a \emph{uniformly accelerated frame}.

\begin{defn}\label{defuasys}
A frame $K'$ is \emph{uniformly accelerated} if there exists a uniformly accelerated family $\{K_\tau(A,\widehat{x}(0),\widehat{\lambda})\}$ such that at every time $\tau$, the frame $K_\tau$ is comoving to $K'$.
\end{defn}
In light of this definition, we may regard our uniformly accelerated observer as positioned at the spatial origin of a uniformly accelerated frame. This approach is motivated by the following statement of Brillouin \cite{Brill}: a frame of reference is a ``heavy
laboratory, built on a rigid body of tremendous mass, as compared to the masses
in motion."

Our construction of a uniformly accelerated family should be contrasted with Mashhoon's approach \cite{Mash3}, which is well suited to curved spacetime, or a manifold setting.
There, the orthonormal basis is defined by
\begin{equation}\label{Mashhoondeflam2}
 c\frac{d\lambda_{(\kappa)}^{\mu}}{d\tau}=\widetilde{A}^{(\nu)}_{(\kappa)}\lambda_{(\nu)}^{\mu},
\end{equation}
where $\widetilde{A}=\widetilde{A}_{\mu\nu}$ is an antisymmetric tensor. Notice that the derivative of each of Mashhoon's basis vectors depends on \emph{all} of the basis vectors, whereas the derivative of each of our basis vectors depends only on its own components. In particular, Mashhoon's observer's four-acceleration depends on both his four-velocity $\lambda_{(0)}$ \emph{and} on the spatial vectors of his basis, while our observers's four-acceleration depends \emph{only} on his four-velocity. This seems to be the more natural physical model: is there any \emph{a priori} reason why the four-acceleration of the observer should depend on his \emph{spatial} basis? We show now, however, that the two approaches are, in fact, equivalent.

The two approaches are equivalent if we identify Mashhoon's tensor $\widetilde{A}$ as our tensor $A$ computed along the worldline: $\widetilde{A}=A(\tau)=A(0)$.
Then, by equation (\ref{uamlam2}) and observation (8) above, we have
\[ c\frac{d\lambda_{(\kappa)}^{\mu}}{d\tau}=A^{\mu}_{\nu}\lambda_{(\kappa)}^{\nu}=\lambda_{(\nu)}^{\mu}\widetilde{A}^{(\nu)}_{(\kappa)} , \]
which is (\ref{Mashhoondeflam2}).

Despite the equivalence of the two approaches, the four-acceleration of Mashhoon's observer \emph{does} depend on his spatial basis. This is because the components of Mashhoon's tensor $\widetilde{A}$ are computed in the comoving frame. In \emph{this} frame, the observer's spatial basis must be perpendicular to his four-velocity. In the inertial lab frame $K$, on the other hand, there is no such restriction.

Unless specifically mentioned otherwise, we will always choose the lab frame $K$ to be the initial comoving frame $K_0$. This implies that $\widehat{\lambda}=I$.  Moreover, \textit{we will always use the acceleration tensor as computed in the initial comoving frame} $K_0$ \textit{ and denote it by} $A$ \textit{instead of by} $\widetilde{A}$.

We now show that all of our solutions of equation (\ref{uam2ivp}) have constant acceleration in the comoving frame. Let $A$ be as in (\ref{aab}). Since $u(0)=(1,0,0,0)^T$, $u(\tau)=\exp(A\tau/c)u(0)$, and $A\lambda=\lambda A$, we have
\begin{equation}\label{aisconstgenproof}
a(\tau)=Au(\tau)=A\exp(A\tau/c)u(0)=
A\lambda(\tau)u(0)=
\lambda(\tau)Au(0)=\lambda_{(i)}(\tau)\mathbf{g}^{(i)}.
\end{equation}
Thus, the acceleration of the observer in the comoving frame is constant and equals $\mathbf{g}$.
\vskip0.3cm

We end this section by showing that if $K'$ and $K''$ are two uniformly accelerated frames with a common acceleration tensor $A$, then the spacetime transformations between $K'$ and $K''$ are \emph{Lorentz}, despite the fact that neither $K'$ nor $K''$ is inertial:
\begin{equation}\label{sameA}
\exp(\widehat{\lambda}A\widehat{\lambda}^{-1}\tau/c)\widehat{\lambda}=\widehat{\lambda}\exp(A\tau/c)\widehat{\lambda}^{-1}\widehat{\lambda}=
\widehat{\lambda}\exp(A\tau/c).
\end{equation}
This implies, in particular, that there is a Lorentz transformation from a lab frame on Earth to an airplane flying at constant velocity, since we are both subject to the same gravitational field.

\section{Spacetime Transformations from a $K'$ to $K$}\label{spacetimetrans}

$\;\;\;$
In this section, we construct the spacetime transformations from a uniformly accelerated frame $K'$ to the inertial frame $K=K_0$. This will be done in two steps.
\vskip0.4cm\noindent
Step 1: From $K_\tau$ to $K$
\vskip0.2cm\noindent
First, we will derive the spacetime transformations from $K_\tau$ to $K$. The idea here is as follows. Let $\widehat{x}(\tau)$ be the worldline of a uniformly accelerated observer. Fix an event $X$. Find the time $\tau$ for which $\widehat{x}(\tau)$ is simultaneous to $X$ in the comoving frame $K_\tau$. Define the $0$-coordinate in $K_\tau$ to be $y^{(0)}=c\tau$. Use the basis $\lambda(\tau)$ of $K_\tau$ to write the relative spatial displacement of the event $X$ with respect to the observer as $y^{(i)}\lambda_{(i)}(\tau),i=1,2,3$. The spacetime transformation from $K_\tau$ to $K$ is then defined to be
\begin{equation}\label{sttransexp}
x^\mu=\widehat{x}^\mu(\tau)+y^{(i)}\lambda_{(i)}^\mu(\tau).
\end{equation}

Transformations of the form (\ref{sttransexp}) have a natural physical interpretation: the vector sum of the \emph{observer} positioned at the origin of the comoving frame $K_\tau$ and the \emph{spatial coordinates} of the event as measured in $K_\tau$. These transformations were also used in \cite{Mash3}. A similar construction can be found in \cite{al1}, in which the authors use \emph{radar 4-coordinates}, and in \cite{MTW}, but in the less general setting of (non-rotating) Fermi-Walker transport.

The above construction relies on the splitting of spacetime into \emph{locally} disjoint 3D spatial hyperplanes $X_{\tau}$. This is indeed possible. Since $X_{\tau}$
is \emph{perpendicular} to $u(\tau)$, there exist a neighborhood of $\tau$ and a spatial neighborhood of the observer in which the $X_{\tau}$ are
pairwise disjoint. This insures that, at least locally, the same event does not occur at two different times. Hence, the observer may uniquely define coordinates for any event within the locality restriction. Thus, at least \emph{locally}, the spacetime transformations from $K_\tau$ to $K$ are given by (\ref{sttransexp}).

Note then when $K'$ is inertial ($A=0$), one may split spacetime into \emph{globally} pairwise disjoint 3D spatial hyperplanes. In this case, there are \emph{no} locality restrictions, and the transformations (\ref{sttransexp}) are defined \emph{everywhere}. We show now, in fact, that the transformations (\ref{sttransexp}) extend the \emph{Lorentz transformations}.

The Lorentz transformations are normally written
\begin{equation}\label{NormalLT}
x^\mu=\Lambda^\mu_\nu y^\nu, \quad x^\mu=(x^0=ct,x^1,x^2,x^3),
\end{equation}
where the $4\times 4$ matrix $\Lambda$ has \emph{constant} entries. Note, however, that the \emph{columns} $\lambda_{(0)},\lambda_{(1)},\lambda_{(2)},\lambda_{(3)}$ of $\Lambda$
form an orthonormal basis of Minkowski space and that $\lambda_{(0)}=u$, the four-velocity of the observer. Rewriting (\ref{NormalLT}) in terms of this basis, we obtain
\begin{equation}\label{NewLT}
x^\mu=y^{(\nu)}\lambda^\mu_{(\nu)}.
\end{equation}
Separating out the time component, we arrive at
\begin{equation}\label{NewLTsplit}
x=y^{(0)}\lambda_{(0)}+y^{(i)}\lambda_{(i)},
\end{equation}
showing that the coordinates of an event in the lab frame $K$ are the usual vector sum of the \emph{observer} positioned at the origin of the comoving frame $K_\tau$ and the \emph{spatial coordinates} of the event as measured in $K_\tau$.

\vskip0.4cm\noindent
Step 2: From $K'$ to $K_\tau$
\vskip0.2cm\noindent
At this point, we invoke a weaker form of the Hypothesis of Locality introduced by
Mashhoon \cite{Mash1,Mash2}. This \emph{Weak Hypothesis of
Locality} is an extension of the Clock Hypothesis.
\vskip0.2cm \textbf{The Weak Hypothesis of
Locality}\quad \emph{Let $K'$ be a uniformly accelerated frame, with an accelerated
observer with worldline $\widehat{x}(\tau)$. For any time $\tau_0$, the rates of the clock of
the accelerated observer and the clock at the origin of the comoving frame $K_{\tau_0}$ are the
same, and, for events simultaneous to $\widehat{x}(\tau_0)$ in the comoving frame $K_{\tau_0}$, the comoving and the accelerated observers measure the same spatial components.}
\vskip0.2cm
Consider an event $X$. By step 1, the coordinates of $X$ in $K$ are $x=\widehat{x}(\tau_0)+y^{(i)}\lambda_{(i)}(\tau_0)$, where $\tau_0$ is the unique value of $\tau$ such that
$X$ and $\widehat{x}(\tau_0)$ are simultaneous in the comoving frame, and $(y^{(0)}=c\tau_0,y^{(i)})$ are the coordinates of $X$ in $K_{\tau_0}$.
Since $x$ and $\widehat{x}(\tau_0)$ are simultaneous in $K_{\tau_0}$, the Weak Hypothesis of Locality implies that the spatial
coordinates $y^{(i)}$ coincide with the spatial coordinates in $K'$. Therefore,
\textit{the spacetime transformations from $K'$ to $K$} are
\begin{equation}\label{xy}
x=\widehat{x}(\tau)+y^{(i)}\lambda_{(i)}(\tau), \;\;\mbox{ with}\;\; \tau=y^{(0)}/c.
\end{equation}

We end this section by calculating the metric at the point $y$ of $K'$. First, we calculate the
differential of the transformation (\ref{xy}). Differentiating (\ref{xy}), we have
\[dx= \lambda_{(0)}(\tau)
dy^{(0)}+\lambda_{(i)}(\tau)dy^{(i)}+
y^{(i)}\frac{1}{c}\frac{d\lambda_{(i)}}{d\tau}dy^{(0)}.\]
Define $\bar{y}=(0,\mathbf{y})$. Using (\ref{Mashhoondeflam2}) (but writing $A$ for $\widetilde{A}$, as is our convention), this becomes
\begin{equation}\label{differential gen2}
dx=\lambda_{(0)}(\tau)
dy^{(0)}+\lambda_{(i)}(\tau)dy^{(i)}+c^{-2}(A\bar{y})^{(\nu)}\lambda_{(\nu)}(\tau)dy^{(0)}\,.
\end{equation}
Finally, since
\begin{equation}\label{Aactson4vector1}
A\bar{y}=(\mathbf{g}\cdot \mathbf{y},\mathbf{y}\times c \boldsymbol{\omega}),
\end{equation}
we obtain
\begin{equation}\label{differential gen1}
dx=\left(\left(1+\frac{\mathbf{g}\cdot\mathbf{y}}{c^2}\right)\lambda_{(0)}+c^{-1}(\mathbf{y}\times
\boldsymbol{\omega})^{(i)}\lambda_{(i)}\right)dy^{(0)}+\lambda_{(j)}dy^{(j)}.
\end{equation}
Therefore, the metric at the point $\bar{y}$ is
\begin{equation}\label{metricaty}
s^2=dx^2=\left( \left(1+\frac{\mathbf{g}\cdot\mathbf{y}}{c^2}\right)^2-c^{-2}(\mathbf{y}\times
\boldsymbol{\omega})^2\right)(dy^{(0)})^2+\frac{2}{c}(\mathbf{y}\times
\boldsymbol{\omega})_{(i)}dy^{(0)}dy^{(i)}+\delta_{jk}dy^{(j)}dy^{(k)}.
\end{equation}
This formula was also obtained by Mashhoon \cite{Mash3}. We point out that the metric is dependent only on the \emph{position} in the accelerated frame and not on \emph{time}.

\section{Examples of Uniformly Accelerated frames and Spacetime Transformations}\label{examples}

$\;\;\;$
In this section, we consider examples of uniformly accelerated frames and the corresponding spacetime transformations. \vskip0.5cm\noindent

\subsection{Null Acceleration\quad ($\alpha=0,\beta=0$)}

Since, in this case, $|\mathbf{g}|=|c\boldsymbol{\omega}|$ and $\mathbf{g} \cdot \boldsymbol{\omega}=0$, we may choose $\mathbf{g}=(g,0,0)$ and
$c\boldsymbol{\omega}=(0,0,g)$.
From (\ref{aab}), we have
\begin{equation}\label{Atype0ex1}
A^{\mu}_{\nu}=\left(\begin{array}{cccc}0 & g & 0 &0\\g&0&g&0 \\
0&-g&0&0\\0&0&0&0\end{array}\right).
\end{equation}
Then
\begin{equation}\label{Asqtype0ex1}
A^2=\left(\begin{array}{cccc}g^2 & 0 & g^2 & 0\\0&0&0&0 \\
-g^2&0&-g^2&0\\0 & 0 & 0 &0\end{array}\right).
\end{equation}
Thus, from (\ref{gensolncmf}), we have
\begin{equation}\label{lambdatype0ex1}
\lambda(\tau)=I+A\tau/c+\frac{1}{2}A^2\tau^2/c^2=\left(\begin{array}{cccc}
1+\frac{g^2\tau^2}{2c^2} & g\tau/c & \frac{g^2\tau^2}{2c^2} & 0 \\ g\tau/c & 1 & g\tau/c & 0 \\
-\frac{g^2\tau^2}{2c^2} & -g\tau/c & 1-\frac{g^2\tau^2}{2c^2} & 0 \\ 0 & 0 & 0 & 1
\end{array}\right).
\end{equation}
The observer's four-velocity is, therefore,
\begin{equation}\label{lambda0type0ex1}
u(\tau)=\lambda_{(0)}(\tau)=\left(1+\frac{g^2\tau^2}{2c^2},g\tau/c,-\frac{g^2\tau^2}{2c^2},0 \right).
\end{equation}
His four-acceleration is
\begin{equation}\label{acceltype0ex1}
a(\tau)= \left(\frac{g^2\tau}{c},g,-\frac{g^2\tau}{c},0 \right)=g\lambda_{(1)}(\tau),
\end{equation}
which shows that the acceleration is constant in the comoving frame.

Integrating (\ref{lambda0type0ex1}), we have
\[ \widehat{x}(\tau)=    \left( c\tau+\frac{g^2\tau^3}{6c},\frac{g\tau^2}{2},-\frac{g^2\tau^3}{6c},0\right). \]
Using (\ref{lambdatype0ex1}) and $y^{(0)}=c\tau$, the spacetime transformations (\ref{xy}) are
\begin{equation}\label{stttype0}
\left(\begin{array}{c}x^0 \\  \; \\ x^1 \\ \; \\ x^2 \\\; \\ x^3 \end{array}\right)
 =\left( \begin{array}{c}c\tau+ \frac{g^2\tau^3}{6c}+y^{(1)}g\tau/c +y^{(2)}\frac{g^2\tau^2}{2c^2} \\ \; \\ \frac{g\tau^2}{2}+y^{(1)}+y^{(2)}g\tau/c \\ \; \\
  -\frac{g^2\tau^3}{6c}- y^{(1)}g\tau/c+y^{(2)}-y^{(2)}\frac{g^2\tau^2}{2c^2} \\ \; \\ y^{(3)} \end{array} \right).
\end{equation}

\subsection{Linear Acceleration\quad ($\alpha>0,\beta=0$)}

Without loss of generality, we may choose
\begin{equation}\label{Atype1ex1}
A^{\mu}_{\nu}= \left(\begin{array}{cccc}0 & g & 0 &0\\g&0&c\omega & 0 \\
0&-c\omega&0&0\\0&0&0&0\end{array}\right),
\end{equation}
where $g>c\omega>0$. In order to simplify the calculation of the exponent of $A$, we perform a Lorentz boost
\begin{equation}\label{lineardriftboost}
 B= \left(\begin{array}{cccc}  g/\alpha & 0 & -c\omega/\alpha& 0  \\
0 & 1 & 0 & 0 \\
-c\omega/\alpha & 0 & g/\alpha & 0 \\
 0 & 0 & 0 & 1
 \end{array}\right)
\end{equation}
to the drift frame corresponding to the velocity
\begin{equation}\label{driftvelocitylin}
\mathbf{v} =(c^2/g,0,0)\times (0,0,\omega).
\end{equation}
Since $g>c\omega>0$, we have $|\mathbf{v}|\le c$.

In the drift frame, the acceleration tensor $A$ becomes
\[ A_{dr}=B^{-1}AB=\left(\begin{array}{cccc}  0 & \alpha & 0 & 0  \\
\alpha & 0 & 0 & 0 \\
0 & 0 & 0 & 0\\
 0 & 0 & 0 & 0
 \end{array}\right),\]
and leads to 1D hyperbolic motion.
Hence,
\[ \lambda(\tau)=\exp(A\tau/c)=B\exp(A_{dr}\tau/c)B^{-1}\]
\begin{equation}\label{lambdatype1ex1}
= \left(\begin{array}{cccc}\frac{g^2}{\alpha^2}\left(\cosh(\alpha\tau/c)-1\right)+1 &
\frac{g}{\alpha}\sinh(\alpha\tau/c)
 & \frac{gc\omega}{\alpha^2}\left(\cosh(\alpha\tau/c)-1\right) & 0  \\ \frac{g}{\alpha}\sinh(\alpha\tau/c)&\cosh(\alpha\tau/c)&\frac{c\omega}{\alpha}\sinh(\alpha\tau/c)&
 0\\
\frac{-gc\omega}{\alpha^2}\left(\cosh(\alpha\tau/c)-1\right)&\frac{-c\omega}{\alpha}\sinh(\alpha\tau/c)&
\frac{-c^2\omega^2}{\alpha^2}\left(\cosh(\alpha\tau/c)-1\right)+1&0\\
0&0&0&1\end{array}\right).
\end{equation}
If $\omega=0$, we recover the usual hyperbolic motion of a frame. Thus, the previous formula is a covariant extension of hyperbolic motion.

From the first column of (\ref{lambdatype1ex1}), the observer's four-velocity is
\begin{equation}\label{4veltype1ex1}
u(\tau)= (\frac{g^2}{\alpha^2}\left(\cosh(\alpha\tau/c)-1\right)+1,\frac{g}{\alpha}\sinh(\alpha\tau/c),\frac{-gc\omega}{\alpha^2}\left(\cosh(\alpha\tau/c)-1\right),0)  .
\end{equation}
Hence, the observer's four-acceleration is
\begin{equation}\label{acceltype1ex1}
a(\tau)=  \left(\frac{g^2}{\alpha}\sinh(\alpha\tau/c),g\cosh(\alpha\tau/c),\frac{-gc\omega}{\alpha}\sinh(\alpha\tau/c),0\right)
=g\lambda_{(1)}(\tau),
\end{equation}
which shows that the acceleration is constant in the comoving frame.

To see how formula (\ref{4veltype1ex1}) extends 1D hyperbolic motion, we compute the proper velocity $\mathbf{u}$. Using (\ref{4veltype1ex1}), we have
\[ \mathbf{u}=\left(\frac{cg}{\alpha}\sinh(\alpha\tau/c),
\frac{-gc^2\omega}{\alpha^2}\left(\cosh(\alpha\tau/c)-1\right),0\right).\]
Since $\frac{d\tau}{dt}=\gamma^{-1}$, and $\gamma$ is the zero component of $u(\tau)$, we have
\begin{equation}\label{ex1dudt}
\frac{d\mathbf{u}}{dt}=\frac{d\mathbf{u}}{d\tau}\frac{d\tau}{dt}=\frac{\left(g\cosh(\alpha\tau/c),
 \frac{-gc\omega}{\alpha} \sinh(\alpha\tau/c) ,0 \right)}{\frac{g^2}{\alpha^2}\left(\cosh(\alpha\tau/c)-1\right)+1}.
\end{equation}
If $\omega=0$, then (\ref{ex1dudt}) reduces to the 1D hyperbolic motion $\frac{d\mathbf{u}}{dt}=\mathbf{g}$, where $\mathbf{g}=(g,0,0)$. If $\omega \neq 0$, then $\frac{d\mathbf{u}}{dt}$ will also depend on $\omega$.
Thus, we have here an explicit example of the fact that $\mathbf{F}=\frac{d\mathbf{p}}{dt}$ if and only if $\boldsymbol{\omega}=0$ (see the end of section \ref{1dhm}).

Integrating (\ref{4veltype1ex1}), we have
\[ \widehat{x}(\tau)=    \left( \frac{c^2}{\alpha^2}\left(\frac{g^2}{\alpha}\sinh(\alpha\tau/c)+c\omega\tau\right),\frac{c^2g}{\alpha^2}
\left(\cosh(\alpha\tau/c)-1\right),\frac{-c^2g\omega}{\alpha^2}\left(\frac{c}{\alpha}\sinh(\alpha\tau/c)-\tau\right),
0\right).\]
Using (\ref{lambdatype1ex1}) and $y^{(0)}=c\tau$, the spacetime transformations (\ref{xy}) are
\begin{equation}\label{stttype1}
\left(\begin{array}{c}x^0 \\  \; \\ x^1 \\ \; \\ x^2 \\\; \\ x^3 \end{array}\right)
 =\left( \begin{array}{c}\frac{c^2}{\alpha^2}\left(\frac{g^2}{\alpha}\sinh(\alpha\tau/c)+c\omega\tau\right)+y^{(1)}\frac{g}{\alpha}\sinh(\alpha\tau/c)
  +y^{(2)}\frac{cg\omega}{\alpha^2}\left(\cosh(\alpha\tau/c)-1\right)\\ \; \\ \frac{c^2g}{\alpha^2}\left(\cosh(\alpha\tau/c)-1\right)
  +y^{(1)}\cosh(\alpha\tau/c)+y^{(2)}\frac{c\omega}{\alpha}\sinh(\alpha\tau/c)  \\ \; \\ \frac{-c^2g\omega}{\alpha^2}\left(\frac{c}{\alpha}\sinh(\alpha\tau/c)-\tau\right)-y^{(1)}\frac{c\omega}{\alpha}\sinh(\alpha\tau/c)
  -y^{(2)}\frac{c^2\omega^2}{\alpha^2}\left(\cosh(\alpha\tau/c)-1\right)+y^{(2)}
    \\ \; \\ y^{(3)}   \end{array} \right).
\end{equation}
\subsection{Rotational Acceleration\quad ($\alpha=0,\beta>0$)}

Without loss of generality, we may choose
\begin{equation}\label{Atype2ex1}
A^{\mu}_{\nu}= \left(\begin{array}{cccc}0 & g & 0 &0\\g&0&c\omega&0\\
0&-c\omega&0&0\\0&0&0&0\end{array}\right),
\end{equation}
where $c\omega>g>0$. In order to simplify the calculation of the exponent of $A$, we perform a Lorentz boost
\[B= \left(\begin{array}{cccc}  c\omega/\beta & 0 & -g/\beta& 0  \\
0 & 1 & 0 & 0 \\
-g/\beta & 0 & c\omega/\beta & 0 \\
 0 & 0 & 0 & 1
 \end{array}\right)\]
to the drift frame corresponding to the velocity
\begin{equation}\label{driftvelocityrot}
\mathbf{v} =(g,0,0)\times (0,0,1/\omega).
\end{equation}
Since $c\omega>g>0$, we have $|\mathbf{v}|\le c$.

In the drift frame, the acceleration tensor $A$ becomes
\[ A_{dr}=B^{-1}AB=\left(\begin{array}{cccc}  0 & 0 & 0 & 0  \\
0 & 0 & \beta & 0 \\
0 & -\beta & 0 & 0\\
 0 & 0 & 0 & 0
 \end{array}\right),\]
and leads to pure rotational motion.
Hence,
\[ \lambda(\tau)=\exp(A\tau/c)=B\exp(A_{dr}\tau/c)B^{-1}\]
\begin{equation}\label{lambdatype2ex1}
= \left(\begin{array}{cccc}\frac{g^2}{\beta^2}(1-\cos(\beta\tau/c))+1 & \frac{g}{\beta}\sin(\beta\tau/c) & \frac{gc\omega}{\beta^2}(1-\cos(\beta\tau/c)) & 0 \\ \frac{g}{\beta}\sin(\beta\tau/c)&\cos(\beta\tau/c)&\frac{c\omega}{\beta}\sin(\beta\tau/c)&0\\ \frac{-gc\omega}{\beta^2}(1-\cos(\beta\tau/c))&-\frac{c\omega}{\beta}\sin(\beta\tau/c)&\frac{-c^2\omega^2}
{\beta^2}(1-\cos(\beta\tau/c))+1&0\\0&0&0&1\end{array}\right).
\end{equation}
If $g=0$, we recover the usual rotation of the basis about the $z$ axis. Thus, the previous formula is a covariant extension of rotational motion.

From the first column of (\ref{lambdatype2ex1}), the observer's four-velocity is
\begin{equation}\label{4veltype2ex1}
u(\tau)=\left(\frac{g^2}{\beta^2}(1-\cos(\beta\tau/c))+1,
 \frac{g}{\beta}\sin(\beta\tau/c),\frac{-gc\omega}{\beta^2}(1-\cos(\beta\tau/c)),0 \right).
\end{equation}
Hence, the observer's four-acceleration is
\begin{equation}\label{acceltype2ex1}
a(\tau)=  \left(\frac{g^2}{\beta}\sin(\beta\tau/c),g\cos(\beta\tau/c),\frac{-gc\omega}{\beta}\sin(\beta\tau/c),0\right)
=g\lambda_{(1)}(\tau),
\end{equation}
which shows that the acceleration is constant in the comoving frame.

Integrating (\ref{4veltype2ex1}), we have
\[ \widehat{x}(\tau)= \left( \frac{-c^2g^2}{\beta^3}\sin(\beta\tau/c)+\left(\frac{g^2}{\beta^2}+1\right)c\tau,
\frac{c^2g}{\beta^2}\left(\cos(\beta\tau/c)-1\right), \frac{c^3g\omega}{\beta^3}\sin(\beta\tau/c)-\frac{c^2g\omega}{\beta^2}\tau,0
\right). \]
Using (\ref{lambdatype2ex1}) and $y^{(0)}=c\tau$, the spacetime transformations (\ref{xy}) are
\begin{equation}\label{stttype2}
\left(\begin{array}{c}x^0 \\  \; \\ x^1 \\ \; \\ x^2 \\\; \\ x^3 \end{array}\right)
 =\left( \begin{array}{c}\frac{-c^2g^2}{\beta^3}\sin(\beta\tau/c)+\left(\frac{g^2}{\beta^2}+1\right)c\tau-\frac{gy^{(1)}}{\beta}\sin(\beta\tau/c)
+ \frac{y^{(2)}cg\omega}{\beta^2}(1-\cos(\beta\tau/c))\\ \; \\ \frac{c^2g}{\beta^2}\left(\cos(\beta\tau/c)-1\right)
+y^{(1)}\cos(\beta\tau/c)-\frac{y^{(2)}c\omega}{\beta}\sin(\beta\tau/c) \\ \; \\ \frac{c^3g\omega}{\beta^3}\sin(\beta\tau/c)-\frac{c^2g\omega}{\beta^2}\tau
 + \frac{y^{(1)}c\omega}{\beta}\sin(\beta\tau/c) - \frac{y^{(2)}c^2\omega^2}{\beta^2}(1-\cos(\beta\tau/c))+ y^{(2)}\\ \; \\ y^{(3)}   \end{array} \right).
\end{equation}

\section{Velocity Transformations}\label{veltransandtd}
$\;\;\;$
In this section, we obtain the transformation of a particle's velocity in a uniformly accelerated frame $K'$ to its four-velocity in the initial comoving inertial frame $K=K_0$.

A particle's four-velocity in $K$ is, by definition, $\frac{dx^\mu}{d\tau_p}$, where $x(\tau_p)$ is the particle's worldline, and $\tau_p$ is the particle's proper time. However, from Special Relativity, it is known that the proper time of a particle depends on its velocity. In addition, it is known that the rate of a clock in an accelerated system also depends on its position, as occurs, for example, for linearly accelerated systems, due to gravitational time dilation. As a result, the quantity $d\tau_p$ depends on both the position and the velocity of the particle, that is, on the state of the particle.

Since we do not yet know the particle's proper time, it is not clear how to calculate the particle's four-velocity in $K$ directly from its velocity in $K'$. To get around this problem, we will differentiate the particle's worldline by a parameter $\tilde{\tau}$ instead of $\tau_p$. For convenience, we will choose
$\tilde{\tau}$ to be a constant multiple of the time. For example, we often choose $\tilde{\tau}=c\tau$. The same technique was used by Horwitz and Piron \cite{offshell}, using the four-momentum instead of the four-velocity, thereby introducing the area known as``off-shell" electrodynamics.

We now introduce the following definition.
\begin{defn}\label{defn4Dvelocity}
Let $x^\mu$ be the worldline of a particle. The particle's \emph{4D velocity with respect to the parameter} $\tilde{\tau}$ is denoted by $\tilde{u}$ and is defined by
\begin{equation}\label{gen4dvelocity}
\tilde{u}^\mu=\frac{dx^\mu}{d\tilde{\tau}}.
\end{equation}
\end{defn}
Note that the 4D velocity has the same direction as the four-velocity. In fact, the particle's four-velocity is
\begin{equation}\label{4veland4dvel}
u=\frac{\tilde{u}}{|\tilde{u}|},
\end{equation}
the normalization of $\tilde{u}$. In the particular case $\tilde{\tau}=ct$, where $t$ is the time in an \emph{inertial} frame, then
\begin{equation}\label{inergamma}
|\tilde{u}|=\left|
\frac{dx^\mu}{cdt}\right|=\left|\left(1,\frac{\mathbf{v}}{c}\right)\right|=\sqrt{1-\frac{\mathbf{v}^2}{c^2}}
=\gamma^{-1}.
\end{equation}

Consider now a moving particle in $K'$. Let $\tilde{w}^{(\nu)}=\frac{dy^{(\nu)}}{dy^{(0)}}$ denote the particle's 4D velocity in $K'$ with respect to $y^{(0)}=c\tau$.
We will calculate the particle's 4D velocity in $K$ at the point $\mathbf{y}$ of $K'$, with respect to $c\tau=c\gamma^{-1}t$, where $\gamma$
corresponds to the observer's velocity in $K$.  Recall from formula (\ref{differential gen2}) that the differential of the spacetime transformations
\begin{equation}\label{lta01}
x=\widehat{x}(\tau)+y^{(i)}\lambda_{(i)}
\end{equation}
is
\begin{equation}\label{differential gen3}
dx=\lambda_{(0)}(\tau)
dy^{(0)}+\lambda_{(i)}(\tau)dy^{(i)}+c^{-2}(A\bar{y})^{(\nu)}\lambda_{(\nu)}(\tau)dy^{(0)}\,.
\end{equation}
Thus, the particle's 4D velocity in $K$ at the point $y$ is
\begin{equation}\label{4velgen1}
 \tilde{u}=\frac{1}{c}\frac{dx}{d\tau}=\frac{dx}{dy^{(0)}} =\frac{dy^{(\nu)}}{dy^{(0)}}\lambda_{(\nu)}(\tau)+c^{-2}(A\bar{y})^{(\nu)}\lambda_{(\nu)}(\tau) \,
= ( \tilde{w}^{(\nu)}+
c^{-2}(A\bar{y})^{(\nu)})\lambda_{(\nu)}(\tau) \,,
\end{equation}
where $\bar{y}=(0,\mathbf{y})$.
Thus, the 4D velocity $\tilde{u}$ is the sum of the particle's 4D velocity $\tilde{w}$ within $K'$
and an additional \textit{4D velocity due to the acceleration of} $K'$ with respect to the inertial frame $K$, which we denote
\begin{equation}\label{4dveld2accel}
\tilde{u}_a=c^{-2}(A\bar{y})^{(\nu)}\lambda_{(\nu)}(\tau).
\end{equation}
Finally, from (\ref{4veland4dvel}), the four-velocity of the particle in $K$ is
\begin{equation}\label{time dilation gen}
u=\frac{\tilde{u}}{|\tilde{u}|}=\frac{\tilde{w}+\tilde{u}_a}{|\tilde{w}+\tilde{u}_a|}=\frac{(
\tilde{w}^{(\nu)}+
c^{-2}(A\bar{y})^{(\nu)})\lambda_{(\nu)}(\tau)}{|\tilde{w}+c^{-2}A\bar{y}|}.
\end{equation}
Writing $\tilde{w}^{(\nu)}=(1,\mathbf{w}/c)$, the four-velocity in the $1+3$ decomposition becomes
\begin{equation}\label{4vel1plus3}
u= \frac{\left(1+\frac{\mathbf{g}\cdot\mathbf{y}}{c^2} ,
c^{-1}(\mathbf{w}+\mathbf{y}\times \boldsymbol{\omega})\right)}{\sqrt{\left(1+\frac{\mathbf{g}\cdot\mathbf{y}}{c^2}\right)^2-\left(\frac{\mathbf{w}+\mathbf{y}\times \boldsymbol{\omega}}{c}\right)^2}}.
\end{equation}
By the Equivalence Principle, we may interpret the zero component of the vector in the numerator as the gravitational time dilation. This factor depends only on the translational acceleration,
which is the only cause of the change in time. The spatial part is the correction due to the particle's velocity in $K'$ and the rotational velocity of $K'$ with respect to $K_\tau$.

We now show that substituting $A=0$ in formula (\ref{4velgen1}) yields the Einstein velocity addition formula for inertial systems. Assume that $K'$ is inertial, say with uniform 3D velocity $\mathbf{v}=(v,0,0)$ with respect to the inertial frame $K$. Suppose a particle has 3D velocity $\mathbf{w}$ in $K'$. We wish to compute $\mathbf{v}\oplus_E\mathbf{w}$, defined to be the particle's 3D velocity in $K$. The particle's 4D velocity in $K'$ with respect to $ct'$, where $t'$ is the time in $K'$, is $\tilde{w}^{(\nu)}=(1,\mathbf{w}/c)$. The comoving frame of $K'$ in this case is
\begin{equation}\label{comovIner1}
 \lambda_{(0)}=\gamma(1,v/c,0,0), \lambda_{(1)}=\gamma(v/c,1,0,0),\lambda_{(2)}=(0,0,1,0),\lambda_{(3)}=(0,0,0,1).
\end{equation}
From (\ref{4velgen1}), we get
\begin{equation}\label{utildeeva}
\tilde{u}=\tilde{w}^{(\nu)}\lambda_{(\nu)}=\left(\gamma\left(1+\frac{w^1v}{c^2}\right),\gamma\left(\frac{v+w^1}{c}\right),w^2/c,w^3/c  \right).
\end{equation}
Dividing $c$ times the spatial part of $\tilde{u}$ by the time component, we obtain the 3D velocity of the particle in $K$ as
\begin{equation}\label{eva}
\mathbf{v}\oplus_E\mathbf{w}=\frac{\left(v+w^1,\gamma^{-1}w^2,\gamma^{-1}w^3  \right)}{1+\frac{w^1v}{c^2}}=\frac{\mathbf{v}+P_{\mathbf{v}}\mathbf{w}+\gamma^{-1}(I-P_{\mathbf{v}})
\mathbf{w}}{1+\frac{w^1v}{c^2}},
\end{equation}
where $P_{\mathbf{v}}\mathbf{w}$ denotes the projection of $\mathbf{w}$ onto $\mathbf{v}$. This is the well-known Einstein velocity addition formula (see \cite{Rindler}, formula (3.7)).

\section{Time Dilation}\label{timedilation}
$\;$

We turn now to time dilation. We will compute the time dilation between the clock of a uniformly accelerated observer located at the origin of $K'$ and the clocks at other positions in $K'$. Let $\tau_p$ be the proper time of a particle positioned at $\mathbf{y}$ in $K'$, with 4D velocity $\tilde{w}$ with respect to $y^{(0)}$. Since the four-velocity (in $K$) of the particle is
\begin{equation}\label{dtauyfromlenu}
u=\frac{1}{c}\frac{dx}{d\tau_p}=\frac{1}{c}\frac{dx}{d\tau}\frac{d\tau}{d\tau_p}=\tilde{u}\frac{d\tau}{d\tau_p},
\end{equation}
equation (\ref{4veland4dvel}) implies that
\begin{equation}\label{dtauyequdtau}
d\tau=\frac{1}{|\tilde{u}|}d\tau_p=\tilde{\gamma}d\tau_p,
\end{equation}
where $\tilde{\gamma}:=\frac{1}{|\tilde{u}|}$ is a function of $\tilde{w}$ and $\mathbf{y}$, or, in short, of the state of the particle. The definition of $\tilde{\gamma}$ for accelerated systems is analogous to the definition of $\gamma$ for inertial systems. In fact, we will see below that if $A=0$, then $\tilde{\gamma}=\gamma$.
The factor $\tilde{\gamma}$ expresses the time dilation between the particle at $\mathbf{y}$ and the observer at the origin of $K'$. To obtain the time dilation of the particle with respect to the inertial frame $K$, one must also multiply
by the time dilation of the observer with respect to $K$, which is the zero component of the observer's four-velocity, explicitly obtained in section \ref{explicitsolns}.

We now express the time dilation (\ref{dtauyequdtau}) in the $1+3$ decomposition. If a particle has velocity $\tilde{w}^{(\nu)}=\frac{dy^{(\nu)}}{dy^{(0)}}=(1,\mathbf{w}/c)$ in $K'$, then the time dilation between the particle and the observer is given by
\begin{equation}\label{gtd2}
   d\tau_p=\sqrt{\left(1+\frac{ {\mathbf{g}}\cdot\mathbf{y}}{c^2}\right)^2-\left(\frac{\mathbf{w}+\mathbf{y}\times {\boldsymbol{\omega}}}{c}\right)^2}\,d\tau\,,\quad \tilde{\gamma}=\frac{1}{\sqrt{\left(1+\frac{ \mathbf{g}\cdot\mathbf{y}}{c^2}\right)^2-\left(\frac{\mathbf{w}+\mathbf{y}\times \boldsymbol{\omega}}{c}\right)^2}}.
\end{equation}
Thus, the time dilation between the particle and the observer in $K'$ combines the gravitational time dilation (via the Equivalence Principle) and an additional time dilation due to the velocity of the particle together with the rotational velocity of the system. The same formula was obtained in \cite{Ni}. Note that the expression underneath the square root must be nonnegative. This limits the admissible values for $\mathbf{y}$ and is a manifestation of the locality of the spacetime transformations from section \ref{spacetimetrans}. The same limitation was obtained by Mashhoon \cite{Mash3}.

If $A=0$, then
\[  \tilde{\gamma}=\frac{1}{\sqrt{1 -\frac{\mathbf{w}^2}{c^2}}}=\gamma(\mathbf{w}),\]
expressing the time dilation due to the velocity of the particle in $K'$, which is an inertial system in this case.

For a clock at rest in $K'$, the particular case $ \boldsymbol{\omega}=0$ gives a time dilation of
$1+\frac{\mathbf{g}\cdot\mathbf{y}}{c^2}$, which is the known formula for gravitational
time dilation. If $\mathbf{g}=0$, the time dilation is
$\sqrt{1-\left(\frac{\mathbf{y}\times \boldsymbol{\omega}}{c}\right)^2}$, which is the time
dilation due to the rotational velocity of a rotating system.

The lower order terms of the expansion of the time dilation of (\ref{gtd2}) are
\begin{equation}\label{tdfo}
1+\frac{ {\mathbf{g}}\cdot\mathbf{y}}{c^2}-\frac{1}{2}\left(\frac{\mathbf{w}+\mathbf{y}\times {\boldsymbol{\omega}}}{c}\right)^2=1+\frac{ {\mathbf{g}}\cdot\mathbf{y}}{c^2}-\frac{1}{2}\frac{(\mathbf{y}\times {\boldsymbol{\omega}})^2}{c^2}-\frac{1}{2}\frac{\mathbf{w}^2}{c^2}
-\frac{\mathbf{w}\cdot(\mathbf{y}\times  {\boldsymbol{\omega}})}{c^2}.
\end{equation}
The second term represents the gravitational time dilation. The third and fourth terms are the transversal Doppler shifts due to the rotation of the system and the velocity of the particle, respectively. The last term is new in the setting of flat Minkowski space but was also obtained recently by Gr{\o}n and Braeck (\cite{GB}, equation (29)) in Schwarzschild spacetime.

We now obtain the physical meaning of $\boldsymbol{\omega}$ in the acceleration matrix $A$. From (\ref{4vel1plus3}), the four-velocity of a rest point $\mathbf{y}$ in the comoving frame is
\begin{equation}\label{4vel1plus3rp}
u= \frac{\left(1+\frac{\mathbf{g}\cdot\mathbf{y}}{c^2} ,
c^{-1}(\mathbf{y}\times \boldsymbol{\omega})\right)}{\sqrt{\left(1+\frac{\mathbf{g}\cdot\mathbf{y}}{c^2}\right)^2-\left(\frac{\mathbf{y}\times \boldsymbol{\omega}}{c}\right)^2}},
\end{equation}
and the corresponding 3D velocity is
\begin{equation}\label{vely3D}
\mathbf{v}=\frac{d\mathbf{x}}{dt}=\frac{\mathbf{y}\times  \boldsymbol{\omega}}{1+\frac{ \mathbf{g}\cdot\mathbf{y}}{c^2}}.
\end{equation}
This formula defines the angular velocity of a uniformly accelerated body. Note that for rest points on the axis of rotation, we have $\frac{d\mathbf{x}}{dt}=0$. Also, if $\mathbf{y}$ belongs to the plane perpendicular to $\mathbf{g}$, then $\frac{d\mathbf{x}}{dt}=\mathbf{y}\times  \boldsymbol{\omega}$, the classical angular velocity. Multiplying both sides of (\ref{vely3D}) by $\gamma$, we obtain
\begin{equation}\label{vely3Dtau}
\mathbf{u}=\frac{d\mathbf{x}}{d\tau}=\tilde{\gamma}(\mathbf{y}\times  \boldsymbol{\omega}).
\end{equation}
The explanation of this formula is as follows. Measure the angular velocity of each point with respect to a common clock, in this case, the clock at the origin. Then, since each point of the rotating object must have the same period, the classical angular velocity must be multiplied by the time dilation between the clock at the origin and the clock at the point in question.

\section{The Acceleration of Rest Points in a Uniformly Accelerated Frame}\label{arbobs}
$\;\;\;$
We will show here that every rest point in a uniformly accelerated frame is also uniformly accelerated, but the acceleration differs from point to point.
Until now, our uniformly accelerated observer has been located at the origin of a uniformly accelerated system $K'$. In this section, we consider an observer at rest at an arbitrary point $\mathbf{y}$ of $K'$.

First, we show that this observer is also uniformly accelerated. To see this, note that the coefficients of $\lambda_{(\nu)}(\tau)$ in the four-velocity (\ref{time dilation gen}) do not depend on $\tau$. Hence,
\[ c\frac{du}{d\tau_p}=c\frac{du}{d\tau}\frac{d\tau}{d\tau_p}= c\tilde{\gamma}\frac{du}{d\tau}
 =  c\tilde{\gamma}  u^{(\nu)} \frac{d\lambda_{(\nu)}(\tau)}{d\tau}\]
\begin{equation}\label{gammaAu}
  =\tilde{\gamma}  u^{(\nu)}A\lambda_{(\nu)}(\tau)=\tilde{\gamma}Au.
\end{equation}
This shows that the observer at $\mathbf{y}$ is also uniformly accelerated, with acceleration tensor $\tilde{\gamma}A$. In the particular case $\boldsymbol{\omega}=0$, the comoving frame at $\mathbf{y}$ coincides with the comoving frame of the observer at the origin, and
$\tilde{\gamma}=\frac{1}{1+\frac{ \mathbf{g}\cdot\mathbf{y}}{c^2}}$. Thus, in this case, $\mathbf{g}(\mathbf{y})=\tilde{\gamma}\mathbf{g}$. The same formula was obtained by Franklin \cite{Franklin}.

In the general case ($\boldsymbol{\omega}\neq 0$), we now find explicit expressions for the components $\mathbf{g}(\mathbf{y})$ and $\boldsymbol{\omega}(\mathbf{y})$ of the acceleration matrix $A(\mathbf{y})$ of a rest point $\mathbf{y}$. Assume, without loss of generality, that $\boldsymbol{\omega}$ is in the direction of the positive $z$ axis and that $\mathbf{y}$ has spatial coordinates $(r,0,z)$ in the comoving frame.

In order to compute the acceleration matrix $A(\mathbf{y})$ for the observer at $\mathbf{y}$, we need to calculate the comoving frame at $\mathbf{y}$.
By (\ref{4vel1plus3rp}), the 3D velocity $\mathbf{v}$ of the point $\mathbf{y}$ in the comoving frame $\lambda(\tau)$ is
\begin{equation}\label{vely2}
\mathbf{v}= \frac{\omega r (0,-1,0)}{1+\frac{ \mathbf{g}\cdot\mathbf{y}}{c^2}}\quad,\quad  \gamma=\frac{1+\frac{\mathbf{g}\cdot\mathbf{y}}{c^2}}{\sqrt{\left(1+\frac{\mathbf{g}\cdot\mathbf{y}}{c^2}\right)^2-\left(\frac{\omega r}{c}\right)^2}}
= \tilde{\gamma}\left(1+\frac{\mathbf{g}\cdot\mathbf{y}}{c^2}\right).
\end{equation}
Hence, the initial comoving frame at the point $\mathbf{y}$ is $B_v$, where $B_v$ is a Lorentz boost in the direction of $\mathbf{v}$, with $v=|\mathbf{v}|=\frac{\omega r}{1+\frac{ \mathbf{g}\cdot\mathbf{y}}{c^2}}$. Since the acceleration matrix for this observer is
\begin{equation}\label{accelmatrix}
A(\mathbf{y})=\tilde{\gamma}B_vAB_v^{-1},
\end{equation}
the comoving frame at $\mathbf{y}$ is
\[  \lambda'(\tau')=\exp(A(\mathbf{y})\tau'/c)B_v
 = \exp(\tilde{\gamma}B_vAB_v^{-1}\tau'/c)B_v  = B_v\exp(\tilde{\gamma}A\tau'/c) \]
 \[ = B_v \exp(A\tau/c) = B_v\lambda (\tau).  \]
Hence, from (\ref{accelmatrix}), we obtain
\begin{equation}\label{Aaty}
A(\mathbf{y})=\tilde{\gamma}\left( \begin{array}{cccc}
0  & \gamma g^1-\tilde{\gamma}\omega^2r   & g^2  & g^3\gamma   \\
  \gamma g^1-\tilde{\gamma}\omega^2r   &  0  & \gamma c\omega -\tilde{\gamma}g^1\omega r/c  & 0  \\
  g^2  &  -\gamma c\omega +\tilde{\gamma}g^1\omega r/c   &  0 & \tilde{\gamma}g^3\omega r/c   \\
 g^3\gamma   &  0   & -\tilde{\gamma}g^3\omega r/c   &  0
\end{array} \right).
\end{equation}
For a pure rotation ($\mathbf{g}=0$), we have $\mathbf{g}(\mathbf{y})=(-\gamma^2\omega^2r,0,0)$ and $\boldsymbol{\omega}(\mathbf{y})=
(0,0,\gamma^2\omega)$.

\section{Acceleration Transformations in a Uniformly Accelerated
Frame}\label{acctrans}
$\;\;\;$
Our next goal is to obtain the transformation of a particle's acceleration in $K'$ to its four-acceleration in $K$. First, however, we will calculate the 4D acceleration $\tilde{a}=c\frac{d\tilde{u}}{d\tau}$ in $K$. Let $\tilde{b}=c\frac{d\tilde{w}}{d\tau}$. Using (\ref{4velgen1}), we have
\begin{equation}\label{new4a2}
\tilde{a}=c\frac{d\tilde{u}}{d\tau}=c\frac{d\tilde{w}^{(\nu)}}{d\tau}\lambda_{(\nu)}(\tau)+ {A}\frac{d\bar{y}^{(\nu)}}{dy^{(0)}}\lambda_{(\nu)}(\tau)
+c( \tilde{w}^{(\nu)}+
c^{-2}( {A}\bar{y})^{(\nu)})\frac{d\lambda_{(\nu)}(\tau)}{d\tau}=  \tilde{b}+ {A}\bar{w} + {A}\tilde{u}.
\end{equation}
The quantity $\tilde{d}:=\tilde{b}+ {A}\bar{w}$ is the acceleration of the particle with respect to the comoving frame, since this is the part of $\tilde{a}$ which treats $\lambda(\tau)$ as constant. We now write equation (\ref{new4a2}) in the $1+3$ decomposition. Write $\tilde{b}^{(\nu)}=(0,\mathbf{a}_p)$, where $\mathbf{a}_p$ is the 3D acceleration of the particle in $K'$. Since, for any four-vector $r=(r^0,\mathbf{r})$, we have
\begin{equation}\label{Aactson4vector2}
Ar=(\mathbf{g}\cdot \mathbf{r},r^0\mathbf{g}+\mathbf{r}\times c \boldsymbol{\omega}),
\end{equation}
we obtain
\begin{equation}\label{tildea1plus3}
\tilde{a}^{(\nu)}=\left(2 {\mathbf{g}}\cdot\mathbf{w}+\frac{1}{c} {\mathbf{g}}\cdot (\mathbf{y}\times {\boldsymbol{\omega}}),\mathbf{a}_p+\left(1+\frac{ {\mathbf{g}}\cdot\mathbf{y}}{c^2}\right)
 {\mathbf{g}}+2\mathbf{w}\times c  {\boldsymbol{\omega}}+ (\mathbf{y}\times {\boldsymbol{\omega}})\times {\boldsymbol{\omega}}\right).
\end{equation}
For a rest particle, we have $\tilde{w}^{(\nu)}=(1,0,0,0)$, and, in this case, formula (\ref{tildea1plus3}) becomes
\begin{equation}\label{tildea1plus3rest}
\tilde{a}^{(\nu)}=\left(\frac{1}{c} {\mathbf{g}}\cdot (\mathbf{y}\times {\boldsymbol{\omega}}),\left(1+\frac{ {\mathbf{g}}\cdot\mathbf{y}}{c^2}\right)
 {\mathbf{g}}+ (\mathbf{y}\times  {\boldsymbol{\omega}})\times {\boldsymbol{\omega}}\right).
\end{equation}

Now we obtain the particle's four-acceleration in $K$:
\begin{equation}\label{4accfrom4daccel}
a=c\frac{du}{d\tau_p}=c\frac{du}{d\tau}\frac{d\tau}{d\tau_p}=c\tilde{\gamma}
\frac{d}{d\tau}\left(\tilde{\gamma}\tilde{u}\right)=\tilde{\gamma}^2\tilde{a}-\tilde{\gamma}^2(\tilde{a}\cdot u)u.
\end{equation}
Notice that
\[  A\tilde{u}\cdot u =  A\tilde{u}\cdot \tilde{\gamma}\tilde{u}=0,\]
since $A$ is antisymmetric. Now, substituting $\tilde{a}=\tilde{d}+ {A}\tilde{u}$ into (\ref{4accfrom4daccel}), we have
\begin{equation}\label{4accfrom4daccel2}
a=\tilde{\gamma}^2  {A}\tilde{u}+\tilde{\gamma}^2 \tilde{d}-\tilde{\gamma}^2 (\tilde{d}\cdot u)u
=\tilde{\gamma} \left( {A}u+\tilde{\gamma}( \tilde{d}- (\tilde{d}\cdot u)u)\right) .
\end{equation}
Let $P_u\tilde{d}$ be the projection of $\tilde{d}$ onto $u$, and let $\tilde{d}_{\perp}=(I-P_u)\tilde{d}$.
Then we can write the four-acceleration as
\begin{equation}\label{4accperp2}
a=\tilde{\gamma}^2 ( {A}\tilde{u}+\tilde{d}_\perp)=\tilde{\gamma} {A}u+
\tilde{\gamma}^2\tilde{d}_\perp.
\end{equation}

The explanation of formula (\ref{4accperp2}) is as follows. The acceleration of a rest point in the comoving frame is $ {A}u$, where $u$ is the point's four-velocity in $K$. The factor $\tilde{\gamma}$ is the time dilation factor between the observer's clock and the proper time of the particle and arises here because we differentiated the four-velocity by $\tau$ instead of $\tau_p$.
The term $\tilde{\gamma} {A}u$ accounts only for the acceleration of $K'$ with respect to $K$. Thus, we must add the term
$\tilde{\gamma}^2\tilde{d}_\perp$ to account for the acceleration of the particle inside $K'$. In this term, the factor $\tilde{\gamma}^2$ appears because we differentiated
\emph{twice} by $\tau$ instead of $\tau_p$. Since the four-acceleration is always perpendicular to the four-velocity, and $ {A}u$ is perpendicular to $u$, the four-acceleration can contain only the component of $\tilde{d}$ which is perpendicular to $u$. This completes the explanation of formula (\ref{4accperp2}).

Consider the motion of a charged particle in a constant electromagnetic field $F$. We decompose its motion into motion under a constant Lorentz force and acceleration produced by the self-force due to the radiation.  We consider the particle to be at the origin ($\tilde{\gamma}=1$) of a uniformly accelerated system $K'$, with acceleration tensor $A=\frac{e}{m}F$. The acceleration due to the radiation will be considered as motion with respect to $K'$.  The self-force generates an acceleration, which is known to be $\tilde{d}=\tau_0 A^2u$, where $\tau_0$ is a universal constant. Thus, in this case, formula (\ref{4accfrom4daccel2}) coincides with the  Lorentz-Abraham-Dirac equation (\cite{Rohrlich2}, equation S-10, page 259)
 \begin{equation}\label{rohreq}
    \frac{du}{d\tau}=Au + \tau_0\left(A^2u - (A^2 u \cdot u)u\right),
 \end{equation}
which Rohrlich calls the \emph{correct} equation of motion of a classical point charge.

If the acceleration due to the radiation is in the direction of the four-velocity, then $\tilde{d}_{\perp}=0$. In this case, the motion of a uniformly accelerated charge also satisfies equation (\ref{rohreq}). This occurs, for example, in the particular cases of 1D hyperbolic motion ($\boldsymbol{\omega}=0$) and pure rotation ($\mathbf{g}=0$).


\section{Summary and Discussion}\label{conc}
$\;\;\;$
The fully Lorentz covariant \emph{Relativistic Dynamics Equation} (\ref{uam2}) extends the 3D relativistic dynamic equation $\mathbf{F}=\frac{d\mathbf{p}}{dt}$. We have shown that the \emph{standard} 4D equation $F=\frac{dp}{d\tau}$ is only partially covariant. To achieve full Lorentz covariance, we replaced the four-force $F$ by a rank 2 antisymmetric \emph{tensor} $A_{\mu\nu}$ acting on the four-velocity.

In section \ref{explicitsolns}, we obtained \emph{explicit} solutions to (\ref{uam2}) in the case of \emph{constant force}. We call the solutions \emph{uniformly accelerated motion}. We have shown that uniformly accelerated motions are divided into four \emph{Lorentz-invariant} types: null, linear, rotational, and general.  For \emph{null} acceleration, the worldline (\ref{xtype0}) is \emph{cubic} in the time. \emph{Linear} acceleration (\ref{utype1}) covariantly extends 1D hyperbolic motion, while \emph{rotational} acceleration (\ref{utype2}) covariantly extends pure rotational motion.
We have shown that if we keep the tensor $A$ constant, the nonrelativistic limit (\ref{classicallimitlinear2}) of our uniformly accelerated motion is motion with constant linear acceleration. A different nonrelativistic limit is obtained for rotational and general uniform acceleration by keeping the components $\mathbf{g}$ and $\boldsymbol{\omega}$ of the tensor $A$ constant. This limit (\ref{nrl25}) describes motion under a Lorentz-type force, which includes uniform rotation.

In \cite{ProcYF}, it is shown that a \emph{photon} is uniformly accelerated. To which of the four types of uniform acceleration does the photon belong? The two Lorentz invariants of the electromagnetic field of a photon are both $0$, as in a constant, uniform electromagnetic field in the absence of sources. Since motion in such a field is an example of \emph{null} acceleration, we conjecture that the motion of a photon is described by null acceleration.

Our results here are restricted to the particular case of \emph{constant force}. In \cite{Fho}, the first author considers the one-dimensional \emph{non-constant force} case $F=-kx$.

In \cite{FG10}and \cite{ProcYF}, the first author proved that the spacetime transformations between two frames which are linearly uniformly accelerated with respect to each other must be one of two types. Type I assumes Mashhoon's Weak Hypothesis of Locality. The transformations of section \ref{examples} are explicit examples of Type I. If the Weak Hypothesis of Locality fails, then there exists a universal maximal acceleration $a_{max}$, and the resulting transformations are Type II. It still remains to compute the explicit Type II transformations corresponding to those of section \ref{examples}. Since accelerations are bounded by $a_m$, the set of admissible accelerations form a \emph{bounded symmetric domain} known as a $JC^*$-triple. Thus, one could use the machinery of chapter 4 of \cite{F04} to compute Type II transformations.

We have some preliminary results on the \emph{rigidity} of a uniformly accelerated frame, including a possible resolution of the Ehrenfest paradox. This paradox concerns a rotating disk. Let $r$ be the radius of the disk, measured in the lab frame, when the disk is at rest. When the disk rotates, the radius is always perpendicular to the disk's motion. Thus, an observer in the lab frame will measure the radius of the rotating disk to be $r$. However, the circumference should appear to be Lorentz-contracted by a factor of
$\gamma(\omega r)$, where $\omega$ is the angular velocity of the disk. This implies that the radius of the disk is $r\sqrt{1-\frac{(\omega r)^2}{c^2}}$, which is less than $r$. We plan to present our resolution of this paradox in an upcoming paper.

We are also studying ways in which to incorporate \emph{radiation} into our model.

\end{document}